\documentclass[12pt]{article}
\usepackage{amssymb}
\usepackage{amsfonts}
\usepackage{makeidx}
\usepackage{amsmath}
\usepackage{sw20elba}
\usepackage{hyperref}

\newtheorem{theorem}{Theorem}

\newtheorem{remark}[theorem]{Remark}

\input{tcilatex}
\begin{document}

\title{Relativistic dynamics of accelerating particles derived from field
equations.}
\author{Anatoli Babin and Alexander Figotin \\
University of California at Irvine}
\maketitle

\begin{abstract}
In relativistic mechanics the energy-momentum of a free point mass moving
without acceleration forms a four-vector. Einstein's celebrated energy-mass
relation $E=mc^{2}$ is commonly derived from that fact. By contrast, in
Newtonian mechanics the mass is introduced for an accelerated motion as a
measure of inertia. In this paper we rigorously derive the relativistic
point mechanics and Einstein's energy-mass relation using our recently
introduced neoclassical field theory where a charge is not a point but a
distribution. We show that both the approaches to the definition of mass are
complementary within the framework of our field theory. This theory also
predicts a small difference between the electron rest mass relevant to the
Penning trap experiments and its mass relevant to spectroscopic measurements.
\end{abstract}

\section{Introduction}

The concept of a point particle is at the very heart of Newtonian mechanics.
Regarding this, Einstein said, "Physical events, in Newton's view, are to be
regarded as the motions, governed by fixed laws, of material points in
space. The material point is our only mode of representing reality when
dealing with changes taking place in it" \cite{Einstein IO}. Even in the
quantum mechanics the concept of a point particle continues to be
fundamentally exact. As Feynman puts it, "The wave function $\psi \left( 
\mathbf{r}\right) $ for an electron in an atom does not, then, describe a
smeared-out electron with a smooth charge density. The electron is either
here, or there, or somewhere else, but wherever it is, it is a point charge" 
\cite{Feynman III}. \ One of our key motivations for introducing a
neoclassical field theory of distributed elementary charges in \cite{BF4}-%
\cite{BF7} was a desire to account for particle properties as well as for
wave phenomena in a single mathematically sound Lagrangian relativistic
theory. This theory is self-contained and, consequently, all particle
properties must naturally come out from the field equations as
approximations. We have already demonstrated that the theory implies in the
non-relativistic limit (i) the non-relativistic particle mechanics governed
by the Newton equations with the Lorentz forces and (ii) the frequency
spectrum for hydrogenic atoms. In this paper we study relativistic aspects
of the theory, namely, the motion of a localized wave which describes the
distributed charge in an external EM field. \ An idea of a particle emerging
as a well localized field was explored by a number of authors, in particular
in the form of "extended charge" models. The Lorentz-Abraham model and its
developments was studied and advanced in \cite{Appel Kiessling}, \cite%
{BambusiG93}, \cite{ImaikinKM}, \cite{Jackson}, \cite{Kiessling2}, \cite%
{KKunzeSpohn}, \cite{Nodvik}, \cite{Pearle1}, \cite{Rohrlich}, \cite%
{Schwinger}, \cite{Spohn}, \cite{Yaghjian}. In 1905-1906 Poincar\'{e}
suggested \cite{Poincare} (see also \cite{Jackson}, \cite{Rohrlich}, \cite%
{Pauli RT}, \cite{Schwinger}, \cite{Yaghjian} and references within) adding
non-electromagnetic cohesive forces to the Lorentz-Abraham model. Here we
study dynamics in an external field of a distributed charge with such
cohesive forces.

Recall now the fundamentals of the relativistic dynamics of a mass point and
the relativistic field theory. The relativistic dynamics of an accelerating
mass point charge in the case where the acceleration is caused by the
electromagnetic (EM) external field $\mathbf{E}_{\mathrm{ex}}$, $\mathbf{B}_{%
\mathrm{ex}}$ is described (see, for example, \cite{Barut}, \cite{Pauli RT})
by the following equation: 
\begin{equation}
\frac{\mathrm{d}}{\mathrm{d}t}\left( M\mathbf{v}\right) =\mathbf{f}_{\mathrm{%
Lor}}\left( t,\mathbf{r}\right) ,\quad M=m_{0}\gamma ,\quad \gamma =\left( 1-%
\frac{\mathbf{v}^{2}}{\mathrm{c}^{2}}\right) ^{-1/2},  \label{fre1}
\end{equation}%
where $\mathbf{v}=$ $d\mathbf{r/}dt$ is its velocity, $m_{0}$ is the rest
mass of the mass point, $\mathbf{f}_{\mathrm{Lor}}$ is the Lorentz force,%
\begin{equation}
\mathbf{f}_{\mathrm{Lor}}\left( t,\mathbf{r}\right) =q\left( \mathbf{E}_{%
\mathrm{ex}}\left( t,\mathbf{r}\right) +\frac{1}{\mathrm{c}}\mathbf{v}\times 
\mathbf{B}_{\mathrm{ex}}\left( t,\mathbf{r}\right) \right) ,  \label{fLor0}
\end{equation}%
$q$ is its charge and $\gamma $ is known as the Lorentz\textbf{\ }factor.
Equations (\ref{fre1}) for the space components of the 4-vector are usually
complemented with the time component 
\begin{equation}
\frac{\mathrm{d}}{\mathrm{d}t}\left( M\mathrm{c}^{2}\right) =\mathbf{f}_{%
\mathrm{Lor}}\cdot \mathbf{v.}  \label{ftime}
\end{equation}%
In particular, for small velocities when $\left\vert \mathbf{v}\right\vert /%
\mathrm{c}\ll 1$ we readily recover from equation (\ref{fre1}) as its
non-relativistic approximation  Newton's equation with the Lorentz force by
setting $\gamma =1$. Note that in (\ref{fre1}) the rest mass $m_{0}$ of a
point is an intrinsic property of a point and is prescribed.

In a relativistic field theory the relativistic field dynamics is derived
from a relativistic covariant Lagrangian. The field equations, total energy,
momentum, forces and their densities are naturally defined in terms of the
Lagrangian for both closed and non closed (with external forces) systems
(see, for instance, \cite{Anderson}, \cite{Barut}, \cite{Moller}, \cite%
{Pauli RT}, \cite{Sexl}, see also Section \ref{ChTensor}). For a closed
system the total momentum has a simple form $\mathbf{P}=M\mathbf{v}$ with a
constant velocity $\mathbf{v}$, and the relativistic invariance of the
energy-momentum 4-vector allows to derive \ Einstein's celebrated
energy-mass relation,%
\begin{equation}
\mathsf{E}=M\mathrm{c}^{2}\quad M=m_{0}\gamma ,  \label{fre3a}
\end{equation}%
between the mass $M$ and the energy $\mathsf{E}$ (see, for instance, \cite[%
Sec. 7.1-7.5]{Anderson}, \cite[Sec. 3.1-3.3, 3.5]{Moller}, \cite[Sec. 37]%
{Pauli RT}, \cite[Sec. 4.1]{Sexl}). As stated by Pauli, \cite[p. 123]{Pauli
RT}: "We can thus consider it as proved that the relativity principle, in
conjunction with the momentum and energy conservation laws, leads to the
fundamental principle of the equivalence of mass and (any kind of) energy.
We may consider this principle (as was done by Einstein) as the most
important of the results of the theory of special relativity." Yet another
important point about the Einstein mass-energy relations is made by Laue 
\cite[p. 529]{Schilpp}: "... we can determine the total amount of energy in
a body from its mass. We thereby get rid of the arbitrariness of the zero
point of energy which the former definition of energy (cf. Section III) was
forced to introduce. There are not merely energy differences, as before; the
energy possesses a physically meaningful absolute value."

Now let us take a brief look at the difference in determining the mass in
the dynamics of a single point versus a general Lagrangian Lorentz invariant
field theory.

First, since equation (\ref{fre1}) for a single mass point has a form of
Newton's law,\ one can determine as in Newtonian mechanics the mass $M$\ as
a measure of inertia from the known force $\mathbf{f}_{\mathrm{Lor}}$ and
acceleration $\partial _{t}^{2}\mathbf{r}$ (the variability of $\gamma $ can
be ignored for mild accelerations). In a relativistic field theory for a
closed system the energy-momentum is a four-vector, and that allows to
define the total mass and the rest mass of the system in terms of the energy
by Einstein's formula (\ref{fre3a}) in the case of uniform motion. However,
in the case of a general non-closed system (which is the subject of our
primary interest since we study field regimes with acceleration), there is
no canonical way to determine the mass, position, velocity, and
acceleration. For a non-closed system there is even a problem with a sound
definition of the center of mass since "the centre of mass loses its
physical importance" \cite[p. 203]{Moller}. To summarize, in a general\
relativistic field theory the rest mass is defined for a uniform motion,
whereas in the Newtonian mechanics the concept of inertial mass is
introduced through an accelerated motion.

Our manifestly relativistic Lagrangian field theory describes a single
charge by a complex-valued scalar field (wave function) $\psi \left( t,%
\mathbf{x}\right) $ satisfying the nonlinear Klein-Gordon (KG) equation,%
\begin{equation}
-\frac{1}{\mathrm{c}^{2}}\tilde{\partial}_{t}^{2}\psi +\tilde{\nabla}%
^{2}\psi -G^{\prime }\left( \left\vert \psi \right\vert ^{2}\right) \psi -%
\frac{m^{2}\mathrm{c}^{2}}{\chi ^{2}}\psi =0,  \label{KGex}
\end{equation}%
where $m$ is a positive mass parameter, and $\chi $ is a constant which
coincides with (or is close to) the Planck constant $\hbar $. The
expressions for the covariant derivatives in (\ref{KGex}) are 
\begin{equation}
\tilde{\partial}_{t}=\partial _{t}+\frac{\mathrm{i}q}{\chi }\varphi _{%
\mathrm{ex}},\quad \tilde{\nabla}=\nabla -\frac{\mathrm{i}q}{\chi \mathrm{c}}%
\mathbf{A}_{\mathrm{ex}},  \label{dtex}
\end{equation}%
where $q$ is the value of the charge, and $\varphi _{\mathrm{ex}},\mathbf{A}%
_{\mathrm{ex}}$ are the potentials of the\emph{\ external EM field\ }(which
can be thought of as the one produced by all remaining charges of the
original system of many charges, see \cite{BF4}-\cite{BF6} for details). The
nonlinear term $G^{\prime }$ is given by the logarithmic expression%
\begin{equation}
G^{\prime }\left( s\right) =G_{a}^{\prime }\left( s\right) =-a^{-2}\left[
\ln \left( a^{3}s\right) +\ln \pi ^{3/2}+3\right] ,\quad s\geq 0,
\label{paf30}
\end{equation}%
where $a$ is the \emph{charge size parameter}. It is established below that
for certain regimes of accelerated motion the relativistic mass point
equations (\ref{fre1})-(\ref{ftime})\ are an approximation which describes
the behavior of the field $\psi $ when it is well localized. The charge
localization is facilitated by the Poincar\'{e}-like cohesive forces
associated with the nonlinearity $G_{a}^{\prime }\left( s\right) $. In
particular, a free resting charge with the minimal energy and size $a$ has a
Gaussian shape, namely, $\left\vert \psi \right\vert =\pi
^{-3/4}a^{-3/2}e^{-\left\vert x\right\vert ^{2}a^{-2}/2}$. The localization
can be described by the ratio $a/R_{f}$ where $R_{f}$ is a typical length
scale of the spatial variation of the external EM forces.

As it is commonly done, we start with assigning a position to the
distributed charge by using its energy density $u\left( t,\mathbf{x}\right) $
to define its total energy $\mathsf{E}$ and \emph{energy center (or
ergocenter)} $\mathbf{\mathbf{r}}\left( t\right) $ by the formulas 
\begin{equation}
\mathsf{E}\left( t\right) =\int u\left( t,\mathbf{x}\right) \,\mathrm{d}%
\mathbf{x,\quad \mathbf{r}}\left( t\right) =\frac{1}{\mathsf{E}\left(
t\right) }\mathbf{\int \mathbf{x}}u\left( t,\mathbf{x}\right) \,\mathrm{d}%
\mathbf{x.}  \label{enin}
\end{equation}%
If a field $\psi $ satisfies the field equation (\ref{KGex}), its ergocenter 
$\mathbf{\mathbf{r}}$ and energy $\mathsf{E}$ satisfy equations which can be
derived from the conservation laws for the KG equation (\ref{KGex}). We
prove under the assumption of localization in the asymptotic limit $%
a/R_{f}\rightarrow 0$ that these equations turn into the relativistic mass
point equations (\ref{fre1}), (\ref{ftime}). Remarkably, the value of the
inertial mass determined from the equations by the Newtonian approach
coincides exactly with the mass given by Einstein's formula (\ref{fre3a}).
Of course, a convincing argument for the equivalence of the inertial mass
and the energy based on the analysis of the charge momentum when it
interacts with the electromagnetic field has been made by Einstein \cite%
{Einstein05a}, but\ here the same is obtained through a thorough
mathematical analysis of a concrete Lagrangian model.

Since our Lagrangian theory is self-contained, Einstein's energy-mass
formula (\ref{fre3a}) or any asymptotic law of motion must be derived within
the framework of the theory. \emph{Importantly, since the limit mass point
equations are derived, and not postulated, the resulting rest mass is shown
to be an integral of motion rather than a prescribed constant. Consequently,
the rest mass may take different values depending on the state of the field.
In particular, in addition to the primary Gaussian ground state there is a
sequence of rest states with higher rest energies and rest masses.} The
possibility of different rest masses comes from the fact that in our theory
an elementary charge is not a point but is a distribution described by a
wave function $\psi $. The charge although elementary has infinitely many
degrees of freedom, with internal interactions of not electromagnetic
origin, contributing to its internal energy.

Our theory yields a simple expression for the rest mass $m_{0}$ of a charge
which differs slightly from the mass parameter $m$, namely,%
\begin{equation}
m_{0}=m+\frac{m}{2}\frac{a_{\mathrm{C}}^{2}}{a^{2}},  \label{fre3b}
\end{equation}%
where $a_{\mathrm{C}}$ is the reduced Compton wavelength. Evidently, the
difference $m_{0}-m$ vanishes as $a_{\mathrm{C}}/a\rightarrow 0$. In the
nonrelativistic case treated in \cite{BF4}-\cite{BF6} the nonlinear
Schrodinger equation follows from the KG equation in the limit $\mathrm{c}%
\rightarrow \infty $ and $\alpha \rightarrow 0$ where $\alpha =q^{2}\chi
^{-1}\mathrm{c}^{-1}$ is the Sommerfeld fine structure constant. In this
limit the mass parameter $m$ coincides with the inertial mass $m_{0}$ in
evident agreement with the limit of relation (\ref{fre3b}) with a fixed Bohr
radius $a_{\mathrm{B}}=a_{\mathrm{C}}/\alpha $. Observe that our theory
provides for complementary interpretations of the relativistic and
non-relativistic\ masses which are sometimes considered (see \cite%
{EriksenV76}) to be "rival and contradictory."

The focus of this paper is on the charge motion in an external EM field
which has a mild spatial variation. The external EM fields with stronger
spatial variation are covered by the theory as well but with the use of
other techniques. Namely, in the case of the hydrogen atom \cite{BF6}, \cite%
{BF7}, for the electron in the Coulomb field we derive the well known energy
spectrum involving the Rydberg constant $R_{\infty }$ with the mass
parameter $m$ in the place of the electron mass. The difference between the
mass $m$ entering the spectroscopic data and the inertial mass $m_{0}$ as it
appears in the Penning trap experiments is discussed in Section \ref%
{SSpecmass}.

In the following sections we analyze relativistic features of the charge
emerging from the underlying field dynamics. In particular, in Section \ref%
{sreldis} we derive the mass point equations (\ref{fre1}), (\ref{ftime})
under an assumption that the charge wave function remains localized. In
Section \ref{secAR} we show that the localization assumption is consistent
with the KG equation (\ref{KGex}). In particular, we describe there a class
of external EM fields in which the charge maintains its localization in the
strongest possible form when it accelerates according to the KG field
equation (\ref{KGex}). The relativistic theory of many interacting charges
is left for another paper.

\section{Relativistic distributed charge as a particle\label{sreldis}}

In \cite{BF4}-\cite{BF7} we developed a neoclassical theory for many
interacting charges based on a relativistic and gauge invariant Lagrangian.
We have demonstrated there that this Lagrangian theory describes
electromagnetic interaction in all spatial scales: it accounts for at least
some quantum phenomena at atomic scale including the frequency spectrum of
the hydrogen atom, and it accounts for the classical motion of
non-relativistic charges when they are well separated and localized. In this
paper we study the relativistic aspects of our theory in the case of a
single charge. Our general Lagrangian in this special case turns into the
following relativistic and gauge invariant expression 
\begin{equation}
\mathcal{L}_{1}\left( \psi \right) =\frac{\chi ^{2}}{2m}\left\{ \frac{1}{%
\mathrm{c}^{2}}\tilde{\partial}_{t}\psi \tilde{\partial}_{t}^{\ast }\psi
^{\ast }-\tilde{\nabla}\psi \cdot \tilde{\nabla}^{\ast }\psi ^{\ast }-\kappa
_{0}^{2}\psi ^{\ast }\psi -G\left( \psi ^{\ast }\psi \right) \right\} ,
\label{paf1}
\end{equation}%
where $\psi \left( t,\mathbf{x}\right) $ is a complex valued wave function 
over the space-time continuum and $\psi ^{\ast }$ is its complex conjugate.
In the expression (\ref{paf1}) $\mathrm{c}$ is the speed of light, 
\begin{equation}
\kappa _{0}=\frac{m\mathrm{c}}{\chi }.  \label{paf2}
\end{equation}%
The covariant derivatives in (\ref{paf1}) are defined by (\ref{dtex}) and
the nonlinearity $G\left( s\right) $ is defined by the formula 
\begin{equation}
G\left( s\right) =G_{a}\left( s\right) =-a^{-2}s\left[ \ln \left(
a^{3}s\right) +\ln \pi ^{3/2}+2\right] ,\quad s\geq 0,  \label{paf3}
\end{equation}%
where $a>0$ is the size parameter. \ The field equation corresponding to the
Lagrangian (\ref{paf1}) is the nonlinear Klein-Gordon (KG) equation (\ref%
{KGex}).

It is proven in \cite{BF4}-\cite{BF6} that in the \emph{non-relativistic}
case the trajectory of the ergocenter converges\ to a corresponding solution
of Newton's equation with the Lorentz force as $a/R_{f}\rightarrow 0$ where $%
R_{f}$ is a typical variation scale of the external EM field. It becomes our
goal now \emph{to derive from the KG field equation (\ref{KGex} ) the
relativistic law of motion} for the ergocenter $\mathbf{\mathbf{r}}\left(
t\right) $. Namely, we show below that the ergocenter trajectory $\mathbf{%
\mathbf{r}}\left( t\right) $ converges to a solution of the relativistic
equation (\ref{fre1}) as $a/R\rightarrow 0$ if the charge is localized. The
derivation is based entirely on the analysis of the KG equation (\ref{KGex})
and corresponding conservation laws.

\subsection{Field symmetric energy-momentum tensor and conservation laws 
\label{ChTensor}}

There are a few popular conventions in setting up coordinates and the metric
in the Minkowski four dimensional space-time. We pick the one which seems to
be dominant nowadays as in \cite[Section 1]{Barut}, \cite[Sections 1.1-1.4, 2%
]{LandauLif F}, \cite[Section 1-1-1]{Itzykson Zuber}. The time-space
four-vector in its contravariant $x^{\mu }$ and covariant $x_{\mu }$ forms
is represented as follows%
\begin{equation}
x=x^{\mu }=\left( x^{0},x^{1},x^{2},x^{3}\right) =\left( \mathrm{c}t,\mathbf{%
x}\right) ,\ \mu =0,1,2,3;  \label{fre4}
\end{equation}%
\begin{equation}
x_{\mu }=g_{\mu \nu }x^{\nu }=\left( x^{0},-x^{1},-x^{2},-x^{3}\right) ;
\label{fre5}
\end{equation}%
\begin{equation}
\partial _{\mu }=\frac{\partial }{\partial x^{\mu }}=\left( \frac{1}{\mathrm{%
c}}\partial _{t},\nabla \right) ;\ \partial ^{\mu }=\frac{\partial }{%
\partial x_{\mu }}=\left( \frac{1}{\mathrm{c}}\partial _{t},-\nabla \right) ;
\label{fre6}
\end{equation}%
with the common convention on the summation of the same indices. The metric
tensor $\ g_{\mu \nu }=$ $g^{\mu \nu }$ is defined by%
\begin{equation}
\left\{ g_{\mu \nu }\right\} =\left\{ g^{\mu \nu }\right\} =\left[ 
\begin{array}{cccc}
1 & 0 & 0 & 0 \\ 
0 & -1 & 0 & 0 \\ 
0 & 0 & -1 & 0 \\ 
0 & 0 & 0 & -1%
\end{array}%
\right] .  \label{fre7}
\end{equation}%
We also use a common convention for the space 3-vector%
\begin{equation}
x^{i}=\left( x^{1},x^{2},x^{3}\right) =\mathbf{x},\ i=1,2,3,  \label{fre8}
\end{equation}%
emphasizing notationally by the Latin superscript its difference from
4-vector $x^{\mu }$ with the Greek superscript.

For vector field potentials $\left( \varphi ,\mathbf{A}\right) $ in the KG
equation (\ref{KGex}), we use standard relativistic notations with the
four-vector potential $A^{\mu }$, four-vector current density $J^{\nu }$,
and the electromagnetic field $F^{\mu \nu }$: 
\begin{equation}
A^{\mu }=\left( \varphi _{\mathrm{ex}},\mathbf{A}_{\mathrm{ex}}\right)
,\quad J^{\mu }=\left( \mathrm{c}\varrho ,\mathbf{J}\right) ,\quad F^{\mu
\nu }=\partial ^{\mu }A^{\nu }-\partial ^{\nu }A^{\mu },  \label{fre9}
\end{equation}%
so that%
\begin{equation}
F_{\mathrm{ex}}^{\mu \nu }=\left[ 
\begin{array}{cccc}
0 & -E_{1} & -E_{2} & -E_{3} \\ 
E_{1} & 0 & -B_{3} & B_{2} \\ 
E_{2} & B_{3} & 0 & -B_{1} \\ 
E_{3} & -B_{2} & B_{1} & 0%
\end{array}%
\right] ,
\end{equation}%
where, as always, 
\begin{equation}
\mathbf{E}=-\nabla \varphi _{\mathrm{ex}}-\frac{1}{\mathrm{c}}\partial _{t}%
\mathbf{A}_{\mathrm{ex}},\ \mathbf{B}=\nabla \times \mathbf{A}_{\mathrm{ex}}.
\label{fre10}
\end{equation}%
The \emph{Lorentz force density} of the external EM field $F_{\mathrm{ex}%
}^{\mu \nu }$ acting on a 4-current $J^{\mu }$ is of the form 
\begin{equation}
f^{\mu }=\frac{1}{\mathrm{c}}F^{\mu \nu }J_{\nu }=\left( \frac{1}{\mathrm{c}}%
\mathbf{J}\cdot \mathbf{E},\rho \mathbf{E}+\frac{1}{\mathrm{c}}\mathbf{J}%
\times \mathbf{B}\right) .  \label{fre15}
\end{equation}

We turn now to the analysis of the Lagrangian (\ref{paf1}). As a consequence
of its gauge invariance, we obtain the conserved four-current $J^{\nu
}=\left( \mathrm{c}\rho ,\mathbf{J}\right) $ where the charge density and
the current are defined by the following expressions, \cite{BF7},\emph{\ }%
\begin{gather}
\rho =-\left( \frac{\chi q}{m\mathrm{c}^{2}}\func{Im}\frac{\partial _{t}\psi 
}{\psi }+\frac{q^{2}}{m\mathrm{c}^{2}}\varphi _{\mathrm{ex}}\right)
\left\vert \psi \right\vert ^{2},  \label{paf6} \\
\mathbf{J}=\left( \frac{\chi q}{m}\func{Im}\frac{\nabla \psi }{\psi }-\frac{%
q^{2}}{m\mathrm{c}}\mathbf{A}_{\mathrm{ex}}\right) \left\vert \psi
\right\vert ^{2}.  \notag
\end{gather}%
They satisfy the continuity equation%
\begin{equation}
\partial _{t}\rho +\nabla \cdot \mathbf{J}=0  \label{paf6a}
\end{equation}%
implying the charge conservation 
\begin{equation}
\int_{\mathbb{R}^{3}}\rho \left( t,\mathbf{x}\right) \,\mathrm{d}\mathbf{x}=%
\bar{\rho}=q,  \label{rhoq1}
\end{equation}%
where we choose the constant $\bar{\rho}$ to be exactly the charge $q$ as it
arises in Coulomb's law, see \cite{BF4}-\cite{BF7} for details.

There are well understood approaches, due to Belinfante and Rosenfeld, \cite[%
Sec. 32]{LandauLif F}, \cite[Sec. III.4]{Barut}, \cite[Sec. 22]{Lanczos VPM}%
, to constructing the symmetric energy-momentum tensor (EnMT) $T^{\mu \nu }$
from the Lagrangian $\mathcal{L}$ based on its invariance with respect to
the Poincar\'{e} group. The EnMT $T^{\mu \nu }$ corresponding to our
Lagrangian $\mathcal{L}_{1}$ is constructed by the same method as in \cite%
{BF4} and \cite{BF5} yielding as it does there 
\begin{equation}
T^{\mu \nu }=\frac{\chi ^{2}}{2m}\left\{ \left[ \psi ^{;\mu \ast }\psi
^{;\nu }+\psi ^{;\mu }\psi ^{;\nu \ast }\right] -\left[ \psi _{;\mu }^{\ast
}\psi ^{;\mu }-\kappa _{0}^{2}\psi ^{\ast }\psi -G\left( \psi ^{\ast }\psi
\right) \right] g^{\mu \nu }\right\} ,  \label{emtn1}
\end{equation}%
where the covariant derivatives $\psi ^{;\mu }$ are defined by%
\begin{equation}
\psi ^{;\mu }=\left( \partial ^{\mu }+\frac{\mathrm{i}qA_{\mathrm{ex}}^{\mu }%
}{\chi \mathrm{c}}\right) \psi =\left( \frac{1}{\mathrm{c}}\tilde{\partial}%
_{t}\psi ,-\tilde{\nabla}\psi \right) ,\ \tilde{\partial}_{t}=\partial _{t}+%
\frac{\mathrm{i}q\varphi _{\mathrm{ex}}}{\chi },\ \tilde{\nabla}=\nabla -%
\frac{\mathrm{i}q\mathbf{A}_{\mathrm{ex}}}{\chi \mathrm{c}},  \label{emtn2}
\end{equation}%
and $\psi ^{\ast ;\mu }$ is the complex conjugate to $\psi ^{;\mu }$.

We proceed with the interpretation of entries the symmetric EnMT $T^{\mu \nu
}$, \cite[Section 32]{LandauLif F}, \cite[Chapter 3.4]{Morse Feshbach I} 
\begin{equation}
T^{\mu \nu }=T^{\mu \nu }=\left[ 
\begin{array}{cccc}
u & \mathrm{c}p^{1} & \mathrm{c}p^{_{2}} & \mathrm{c}p^{_{3}} \\ 
\mathrm{c}^{-1}s^{1} & -\sigma ^{11} & -\sigma ^{12} & -\sigma ^{13} \\ 
\mathrm{c}^{-1}s^{_{2}} & -\sigma ^{21} & -\sigma ^{22} & -\sigma ^{23} \\ 
\mathrm{c}^{-1}s^{_{3}} & -\sigma ^{31} & -\sigma ^{32} & -\sigma ^{33}%
\end{array}%
\right] ,\ 
\begin{tabular}{|l|l|}
\hline
$u$ & energy density, \\ \hline
$p^{j}$ & momentum density, \\ \hline
$s^{j}=\mathrm{c}^{2}p^{j}$ & energy flux density, \\ \hline
$\sigma ^{ij}=\sigma ^{ji}$ & symmetric stress tensor. \\ \hline
\end{tabular}%
,  \label{paf11}
\end{equation}%
where$\ i,j=1,2,3$.

\begin{remark}
As a consequence of the symmetry of EnMT $T^{\mu \nu }$ we have the
following relation between the field energy flux and the field momentum
densities%
\begin{equation}
\mathbf{s}=\mathrm{c}^{2}\mathbf{p}  \label{paf13}
\end{equation}%
W. Pauli refers to the identity (\ref{paf13}) as a theorem and makes a
comment similar to that of C. Lanczos, \cite[p. 125]{Pauli RT}: "This is the
theorem of the momentum of the energy current, first expressed by Planck$^{%
\text{229}}$ according to which a momentum is associated with each energy
current. This theorem can be considered as an extended version of the
principle of the equivalence of mass and energy. Whereas the principle only
refers to the total energy, the theorem has also something to say on the
localization of momentum and energy."There is an intimate relation between
the concept of particle and the field concept of the \emph{symmetric
energy-momentum tensor} (EnMT) $T^{\mu \nu }$. In particular, the
fundamental Einstein mass-energy relation $\mathsf{E}_{0}=m_{0}\mathrm{c}%
^{2} $ can be interpreted as the symmetry of the energy-momentum tensor, a
point stressed by C. Lanczos, \cite[p. 394]{Lanczos VPM}: " It was Planck in
1909 who pointed out that the field theoretical interpretation of Einstein's
principle can only be the symmetry of the energy-momentum tensor. If the $%
T_{i4}$ ($i=1,2,3$) (i.e. the momentum density) and the $T_{4i}$, the energy
current, did not agree, then the conservation of mass and energy would
follow different laws and the principle $m=E$ could not be maintained. Nor
could a non-symmetric energy-momentum tensor guarantee the law of inertia,
according to which the centre of mass of an isolated system moves in a
straight line with constant velocity."
\end{remark}

The Noether theorem then implies 10 conservation laws: the \emph{energy and
momentum conservation laws}%
\begin{equation}
\partial _{\mu }T^{\mu \nu }=f^{\nu },\quad f^{\nu }=-\frac{\partial 
\mathcal{L}}{\partial x_{\nu }},  \label{paf14}
\end{equation}%
where $f^{\nu }$ is the force density, and the \emph{angular momentum
conservation law}, \cite[Sec. 10.2]{Sexl}, \cite[Sec. II.1, III.4]{Barut},%
\begin{gather}
\partial _{\mu }M^{\mu \nu \gamma }=N^{\nu \gamma },\text{ where}
\label{paf15} \\
M^{\mu \nu \gamma }=x^{\nu }T^{\mu \gamma }-x^{\gamma }T^{\mu \nu },\quad
N^{\nu \gamma }=x^{\nu }f^{\gamma }-f^{\nu }x^{\gamma },  \notag
\end{gather}%
$M^{\mu \nu \gamma }$ is the \emph{angular momentum density tensor,} and $%
N^{\nu \gamma }$ is the \emph{torque density tensor}. Notice that the
angular momentum conservation follows readily from the energy and momentum
conservation law (\ref{paf14}) combined with the symmetry of the EnMT $%
T^{\mu \nu }$.

Using the interpretation (\ref{paf11}) we can recast the energy-momentum
conservation laws (\ref{paf14}) in a more appealing form%
\begin{equation}
\partial _{t}p^{i}=\dsum_{j=1,2,3}\partial _{j}\sigma ^{ji}-f^{i},\quad
f^{i}=\frac{\partial \mathcal{L}}{\partial x_{i}},\ i=1,2,3,  \label{paf16}
\end{equation}%
\begin{equation}
\partial _{t}u=-\dsum_{j=1,2,3}\partial _{j}s^{j}-f^{0},\quad f^{0}=-\frac{%
\partial \mathcal{L}}{\partial t}.  \label{paf17}
\end{equation}%
In the case of a closed system when $f^{i}=0$, the total conserved
quantities are, \cite[(3.76)-(3.77)]{Barut}, \cite[(6), (12), (15b)]{Pauli
RFTh}, \cite[Sec. 10.2]{Sexl}, 
\begin{equation}
P^{\nu }=\dint_{\sigma }T^{\mu \nu }\,\mathrm{d}\sigma _{\mu },\ J^{\nu
\gamma }=\dint_{\sigma }M^{\mu \nu \gamma }\mathrm{d}\sigma _{\mu },
\label{paf15a}
\end{equation}%
where $\sigma $ is any space-like surface, for instance $x_{0}=\limfunc{const%
}$. $P^{\nu }$ is \emph{the} \emph{four-vector of the total energy-momentum}
and $J^{\nu \gamma }=-J^{\gamma \nu }$ is \emph{the total angular momentum
tensor}. Importantly, \emph{for closed systems the conserved total
energy-momentum }$P^{\nu }$\emph{\ and \ angular momentum }$J^{\nu \gamma }$%
\emph{\ transform respectively as 4-vector and 4-tensor} under Lorentz
transformation, and that is directly related to the conservations laws, \cite%
[Section 6.2]{Moller}, \cite[Section 12.10 A]{Jackson}. But \emph{for open
(not closed) systems generally the total energy-momentum }$P^{\nu }$\emph{\
and }$J^{\nu \gamma }$\emph{\ angular momentum do not transform as
respectively 4-vector and 4-tensor}, \cite[Section 7.1, 7.2]{Moller}, \cite[%
Section 12.10 A, 16.4]{Jackson}.

The formula for the 4-microcurrent density $J^{\mu }$ turns into%
\begin{equation*}
J^{\nu }=-\frac{\chi q}{2m}\mathrm{i}\left( \psi \tilde{\partial}^{\nu \ast
}\psi ^{\ast }-\psi ^{\ast }\tilde{\partial}^{\nu }\psi \right) .
\end{equation*}%
Using the same method as in \cite[Sec. 11.7]{BF4} we find that for the
Lagrangian (\ref{paf1}) the general conservation laws (\ref{paf14}) turn into%
\begin{equation}
\partial _{\mu }T^{\mu \nu }=f^{\nu },  \label{emtn2a}
\end{equation}%
where%
\begin{equation}
f^{\nu }=\frac{1}{\mathrm{c}}J_{\mu }F_{\mathrm{ex}}^{\nu \mu }=\left( f^{0},%
\mathbf{f}\right) =\left( \frac{1}{\mathrm{c}}\mathbf{J}\cdot \mathbf{E}_{%
\mathrm{ex}},\rho \mathbf{E}_{\mathrm{ex}}+\frac{1}{\mathrm{c}}\mathbf{J}%
\times \mathbf{B}_{\mathrm{ex}}\right)   \label{emtn2b}
\end{equation}%
with $\mathbf{J,}\rho $ given in (\ref{paf6}). Evidently, $\mathbf{f}$ is
the Lorentz force density for the external field $F_{\mathrm{ex}}^{\nu \mu }$%
. Using the interpretation of the EnMT entries (\ref{paf13}) and (\ref{emtn1}%
) we get the following representations for the energy and the momentum
densities: 
\begin{equation}
u=\frac{\chi ^{2}}{2m}\left[ \frac{1}{\mathrm{c}^{2}}\tilde{\partial}%
_{t}\psi \tilde{\partial}_{t}^{\ast }\psi ^{\ast }+\tilde{\nabla}\psi \tilde{%
\nabla}^{\ast }\psi ^{\ast }+G\left( \psi ^{\ast }\psi \right) +\kappa
_{0}^{2}\psi \psi ^{\ast }\right] ,  \label{emtn3}
\end{equation}%
\begin{equation}
\mathbf{p}=\left( p^{1},p^{2},p^{3}\right) =-\frac{\chi ^{2}}{2m\mathrm{c}%
^{2}}\left( \tilde{\partial}_{t}\psi \tilde{\nabla}^{\ast }\psi ^{\ast }+%
\tilde{\partial}_{t}^{\ast }\psi ^{\ast }\tilde{\nabla}\psi \right) .
\label{emtn4}
\end{equation}

\subsection{ Relativistic dynamics of the energy center of a localized charge%
}

Here we derive equations for the energy center $\mathbf{r}$ based on the
conservation laws introduced in the previous section. Then based on these
equations under asymptotic localization assumptions we derive the
relativistic point mass equations (\ref{fre1}), (\ref{ftime}) for the energy
center $\mathbf{r}$. We set first the total energy, momentum and the force
expressions respectively by%
\begin{equation}
\mathsf{E}\left( t\right) =\int u\left( t,\mathbf{x}\right) \,\mathrm{d}%
\mathbf{x},\quad P^{i}\left( t\right) =\int p^{i}\left( t,\mathbf{x}\right)
\,\mathrm{d}\mathbf{x},\quad F^{i}\left( t\right) =\int f^{i}\left( t,%
\mathbf{x}\right) \,\mathrm{d}\mathbf{x}.  \label{emtn5}
\end{equation}%
The coordinates $r^{i}$ of the energy center $\mathbf{r}$ are given then by 
\begin{equation}
r^{i}=\frac{1}{\mathsf{E}\left( t\right) }\int x^{i}u\left( t,\mathbf{x}%
\right) \,\mathrm{d}\mathbf{x}.  \label{emtn6}
\end{equation}%
The continuity equation (\ref{paf6a}) when multiplied by $\left( \mathbf{x}-%
\mathbf{r}\right) ^{i}$ readily implies the following expression for the
current density $\mathbf{J}$: 
\begin{equation}
\partial _{t}\left( \left( \mathbf{x}-\mathbf{r}\right) ^{i}\rho \right)
+\rho \partial _{t}\mathbf{r}^{i}+\nabla \cdot \left( \left( \mathbf{x}-%
\mathbf{r}\right) ^{i}\mathbf{J}\right) =\mathbf{J}^{i}.  \label{Jexpr}
\end{equation}%
Integrating over the entire space the energy-momentum conservation laws (\ref%
{paf16}) and (\ref{paf17}) we get%
\begin{equation}
\frac{1}{\mathrm{c}}\partial _{t}\mathsf{E}=F^{0},\quad \partial
_{t}P^{i}=F^{i}.  \label{emtn8}
\end{equation}%
For the energy component in (\ref{emtn8}) using (\ref{Jexpr}) we obtain%
\begin{gather}
F^{0}=\int \frac{1}{\mathrm{c}}\mathbf{J}\cdot \mathbf{E}_{\mathrm{ex}}\,%
\mathrm{d}\mathbf{x}=\bar{\rho}\frac{1}{\mathrm{c}}\partial _{t}\mathbf{r}%
\cdot \mathbf{E}_{\mathrm{ex}}\left( t,\mathbf{r}\right) +\delta _{F}^{0},
\label{F0} \\
\delta _{F}^{0}=\frac{1}{\mathrm{c}}\sum_{i}\int \left( \mathbf{E}_{\mathrm{%
ex}}^{i}\partial _{t}\left( \left( \mathbf{x}-\mathbf{r}\right) ^{i}\rho
\right) +\mathbf{E}_{\mathrm{ex}}^{i}\nabla \cdot \left( \left( \mathbf{x}-%
\mathbf{r}\right) ^{i}\mathbf{J}\right) \right) \,\mathrm{d}\mathbf{x+}
\label{delf0} \\
\mathbf{+}\frac{1}{\mathrm{c}}\partial _{t}\mathbf{r}\cdot \int \rho \left( 
\mathbf{E}_{\mathrm{ex}}\left( t,\mathbf{x}\right) -\mathbf{E}_{\mathrm{ex}%
}\left( t,\mathbf{r}\right) \right) \,\mathrm{d}\mathbf{x.}  \notag
\end{gather}%
Integrating the angular momentum conservation law (\ref{paf15}) for $\nu =0$
and $\gamma =i$ over the entire space we obtain%
\begin{equation}
\frac{1}{\mathrm{c}}\partial _{t}\left( \mathrm{c}^{2}tP^{i}-r^{i}\mathsf{E}%
\right) =\mathrm{c}tF^{i}-F^{0}r^{i}-\delta _{f}^{i}  \label{emtn9}
\end{equation}%
where the \emph{particle discrepancy terms} $\delta _{f}^{i}$ are of the form%
\begin{equation}
\delta _{f}^{i}=\int \left( x^{i}-r^{i}\right) f^{0}\,\mathrm{d}\mathbf{x}.
\label{emtn10}
\end{equation}%
The identities (\ref{emtn8}) and (\ref{emtn9}) imply that 
\begin{equation}
P^{i}=\frac{\mathsf{E}}{\mathrm{c}^{2}}\partial _{t}r^{i}-\frac{1}{\mathrm{c}%
}\delta _{f}^{i},\quad i=1,2,3.  \label{emtn11}
\end{equation}%
Now, combining (\ref{emtn11}) with the second equality in (\ref{emtn8}), we
obtain the\ relation \ 
\begin{equation}
\partial _{t}\left( \frac{1}{\mathrm{c}^{2}}\mathsf{E}\partial _{t}\mathbf{r-%
}\frac{1}{\mathrm{c}}\mathbf{\delta }_{f}\right) =\mathbf{F},\quad ,i=1,2,3,
\label{dterdel}
\end{equation}%
with $\mathbf{F}=\left( F^{1},F^{2},F^{3}\right) $ defined by (\ref{emtn5}),
(\ref{emtn2b}). The expressions for $\mathbf{F}$ can be written in the form 
\begin{equation}
\mathbf{F}=\mathbf{f}_{\mathrm{Lor}}\left( t,\mathbf{r}\right) +\mathbf{%
\delta }_{F},  \label{Fdef}
\end{equation}%
where $\mathbf{f}_{\mathrm{Lor}}\left( t,\mathbf{r}\right) $ is the Lorentz
force (\ref{fLor0}) with the remainder term

\begin{gather}
\mathbf{\delta }_{F}=\int (\mathbf{E}_{\mathrm{ex}}\left( t,x\right) -%
\mathbf{E}_{\mathrm{ex}}\left( t,r\right) +\mathrm{c}^{-1}\partial _{t}%
\mathbf{r}^{i}\times \left( \mathbf{B}_{\mathrm{ex}}\left( x\right) -\mathbf{%
B}_{\mathrm{ex}}\left( r\right) \right) )\rho \mathrm{d}\mathbf{x}+
\label{delFdef} \\
+\int \frac{1}{\mathrm{c}}\left( \partial _{t}\left( \left( \mathbf{x}-%
\mathbf{r}\right) \rho \right) +\sum_{l}\partial _{l}\left( \left( \mathbf{x}%
-\mathbf{r}\right) \mathbf{J}^{l}\right) \right) \times \mathbf{B}_{\mathrm{%
ex}}\mathrm{d}\mathbf{x.}  \notag
\end{gather}%
Equations (\ref{F0}) and (\ref{dterdel}) result in the following system of
two\ ergocenter equations: the spatial part 
\begin{equation}
\partial _{t}\left( \frac{\mathsf{E}}{\mathrm{c}^{2}}\partial _{t}\mathbf{%
r-\delta }_{f}\right) =\mathbf{f}_{\mathrm{Lor}}\left( t,\mathbf{r}\right) +%
\mathbf{\delta }_{F},  \label{dterdel1}
\end{equation}%
and the time part%
\begin{equation}
\frac{1}{\mathrm{c}}\partial _{t}\mathsf{E}=\frac{1}{\mathrm{c}}\partial _{t}%
\mathbf{r}\cdot \mathbf{f}_{\mathrm{Lor}}\left( t,\mathbf{r}\right) +\delta
_{F}^{0}.  \label{dtedelf}
\end{equation}%
Observe now that the particle discrepancy terms $\mathbf{\delta }_{f},%
\mathbf{\delta }_{F},\delta _{F}^{0}$ involve in their integrands the
factors $\left( \mathbf{x}-\mathbf{r}\right) $, $\mathbf{B}_{\mathrm{ex}%
}\left( t,\mathbf{x}\right) -\mathbf{B}_{\mathrm{ex}}\left( t,\mathbf{r}%
\right) $, $\mathbf{E}_{\mathrm{ex}}\left( t,\mathbf{x}\right) -\mathbf{E}_{%
\mathrm{ex}}\left( t,\mathbf{r}\right) $ which vanish at the energy center $%
\mathbf{r}$, and they are small in a small neighborhood of $\mathbf{r}$. In
addition to that, the integrands involve the factors $\mathbf{J,}\rho $
which cannot be large outside a small neighborhood of the energy center if
the solution $\psi $ is localized.

Under the assumption that the charge is localized we neglect the particle
discrepancy terms $\mathbf{\delta }_{f},\mathbf{\delta }_{F},\delta _{F}^{0}$
obtaining the following limit system for $\mathbf{r}$ and $\mathsf{E}$: 
\begin{equation}
\partial _{t}\mathsf{E}=\partial _{t}\mathbf{r}\cdot \mathbf{f}_{\mathrm{Lor}%
}\left( t,\mathbf{r}\right) \mathbf{,}  \label{rederg1}
\end{equation}%
\begin{equation}
\partial _{t}\left( \frac{\mathsf{E}}{\mathrm{c}^{2}}\partial _{t}\mathbf{r}%
\right) =\mathbf{f}_{\mathrm{Lor}}\left( t,\mathbf{r}\right) .
\label{rederg2}
\end{equation}%
Equation (\ref{rederg2}) evidently has the form of the relativistic version
of Newton's law of motion with the Lorentz force, namely%
\begin{equation}
\partial _{t}\left( M\partial _{t}\mathbf{r}\right) =\mathbf{f}_{\mathrm{Lor}%
}\left( t,\mathbf{r}\right) ,  \label{Newtrel}
\end{equation}%
provided the mass $M$ is given by Einstein's formula 
\begin{equation}
M=\frac{\mathsf{E}}{\mathrm{c}^{2}}.  \label{EinM}
\end{equation}%
Equation (\ref{rederg1}) has the form of the time-component for the
relativistic point dynamics, see \cite[Section 29, 37]{Pauli RT}, \cite[%
Section II.1]{Barut}. Let us derive now a relation between the inertial mass 
$M$ of moving charge and the rest mass. Using (\ref{rederg1}) we readily
obtain 
\begin{equation}
M\partial _{t}\mathbf{r}\cdot \partial _{t}\left( M\partial _{t}\mathbf{r}%
\right) =M\partial _{t}\mathbf{r}\cdot q\mathbf{E}_{\mathrm{ex}}\left( t,%
\mathbf{r}\right) =M\mathrm{c}^{2}\partial _{t}M,  \label{Mdiff}
\end{equation}%
which implies the relation 
\begin{equation}
M^{2}-\frac{1}{\mathrm{c}^{2}}M^{2}\left( \partial _{t}\mathbf{r}\right)
^{2}=M_{0}^{2},  \label{Mconst}
\end{equation}%
where $M_{0}^{2}$ is the constant of integration. Consequently, we recover
the well-known formula 
\begin{equation}
M=\gamma M_{0},\quad \gamma =\left( 1-\left( \partial _{t}\mathbf{r}\right)
^{2}/\mathrm{c}^{2}\right) ^{-1/2}.  \label{Mgam}
\end{equation}%
The relations (\ref{Newtrel}) and (\ref{Mgam}) readily imply the accelerated
motion equation (\ref{fre1}). As to the equations (\ref{rederg1}) and (\ref%
{rederg2}), they describe asymptotic behavior of the energy and the
ergocenter of the charge when its wave function $\psi $ remains localized in
the course of motion.

\begin{remark}
\label{Rfreemass}We would like to stress again that the rest mass $M_{0}$\
in our treatment is not a prescribed quantity, but it is derived in (\ref%
{Mconst}) as an integral of motion (or, more precisely, an approximate
integral of the field equation which becomes precise in an asymptotic
limit). As any integral of motion, it can take different values for
different "trajectories" of the field. This is demonstrated by the different
values of the rest mass $M_{0}$ for different rest states constructed in the
following Section \ref{snonlin}. The integral of motion $M_{0}$ can be
related to the mass $m_{0}$ of one of resting charges considered in Section %
\ref{Suniform} by the identity%
\begin{equation}
M_{0}=m_{0}  \label{Meqm0}
\end{equation}%
if the velocity vanishes on a time interval or asymptotically as $%
t\rightarrow -\infty $ or $t\rightarrow \infty $. If the velocity $\partial
_{t}\mathbf{r}$ vanishes just at a time instant $t_{0}$ it is, of course,
possible to express the value of $M_{0}$ in terms of $\mathsf{E}=\mathsf{E}%
\left( \psi \right) $ by formulas (\ref{EinM}), (\ref{Mconst}) but the
corresponding $\psi =\psi \left( t_{0}\right) $ may have no relation to the
rest solutions of the field equation with a time independent profile $%
\left\vert \psi \right\vert ^{2}$. It is also possible that $\partial _{t}%
\mathbf{r}$\ never equals zero, and in fact this is a general case since all
three components of velocity may vanish simultaneously only in very special
situations. Hence, there is a possibility of localized regimes where the
value of the "rest mass" $M_{0}$ may differ from the rest mass of a free
charge. In such regimes the value of the rest mass cannot be derived based
on the analysis of the uniform motion as in Section \ref{Suniform}. This
wide variety of possibilities makes even more remarkable the fact that the
inertial mass is well-defined and that the Einstein formula (\ref{fre3a})
holds even in such general regimes where the standard analysis based on the
Lorentz invariance of the uniform motion as in Section \ref{Suniform} does
not apply. In a general case where the localization is not assumed, the
functional%
\begin{equation}
\hat{M}_{0}^{2}=\frac{\mathsf{E}^{2}}{\mathrm{c}^{4}}\left( 1-\frac{1}{%
\mathrm{c}^{2}}\left( \partial _{t}\mathbf{r}\right) ^{2}\right) 
\label{M0gen}
\end{equation}%
extends formula (\ref{Mconst}) to general fields and produces a\
"generalized rest mass" $M_{0}$ defined in terms of the energy and the
ergocenter as follows: 
\begin{equation}
M_{0}^{2}=\lim_{T\rightarrow \infty }\frac{1}{2T}\int_{-T}^{T}\hat{M}%
_{0}^{2}\left( t\right) dt.  \label{Mrestav}
\end{equation}%
The above formula obviously defines the value of the rest mass for a more
general class of field trajectories, and according to (\ref{EinM}) and (\ref%
{Mconst}) produces the value of the integral of motion in the case of
asymptotic localization.
\end{remark}

\begin{remark}
An accelerating particle in an external field is not a closed system, and
there are principal differences between closed and non-closed systems. In
particular, the total momentum and the energy of a closed system are
preserved. For closed systems the particle equation and the momentum
kinematic representation can be derived from the field theory with the use
of the angular momentum conservation, \cite[Sec. 7.1, 72]{Moller}, \cite[%
Sec. 10.2]{Sexl}, \cite[Sec. 23]{Lanczos VPM}. For non-closed systems the
center of energy (also known as center of mass or centroid) and the total
energy-momentum are frame dependent and hence are not 4-vectors, \cite[Sec.
7.1, 7.2]{Moller}, \cite[Sec. 24]{Lanczos VPM}. The rules of the
transformation for the energy-momentum are due to Einstein and Laue, \cite[%
Sec. 43]{Pauli RT}.
\end{remark}

\subsection{Rest states, their energies and frequencies \label{snonlin}}

We suppose for a resting charge $\left\vert \psi \left( t,x\right)
\right\vert $ to be time independent. Such resting states of the charge
exist in the absence of external fields, $\varphi _{\mathrm{ex}}=0$, $%
\mathbf{A}_{\mathrm{ex}}=0$. We are particularly interested in rest states $%
\psi $ that vary harmonically in time and consider solutions to the KG
equation (\ref{KGex}) in the form of a standing wave%
\begin{equation}
\psi =e^{-i\omega t}\breve{\psi}\left( x\right) ,  \label{psiom}
\end{equation}%
where $\breve{\psi}\left( x\right) $ is central-symmetric. The substitution
of (\ref{psiom}) in the KG equation (\ref{KGex}) yields the following \emph{%
nonlinear eigenvalue problem} 
\begin{equation}
\nabla ^{2}\breve{\psi}=G_{a}^{\prime }\left( \left\vert \breve{\psi}%
\right\vert ^{2}\right) \breve{\psi}+\left( \frac{\omega _{0}^{2}}{\mathrm{c}%
^{2}}-\frac{\omega ^{2}}{\mathrm{c}^{2}}\right) \breve{\psi}=0.  \label{eiga}
\end{equation}%
Recall now that the solution $\breve{\psi}$ must also satisfy the charge
normalization condition (\ref{rhoq1}) which takes the form 
\begin{equation}
\int \left\vert \breve{\psi}\right\vert ^{2}\,\mathrm{d}x=\frac{\omega _{0}}{%
\omega },\qquad \omega _{0}=\frac{m\mathrm{c}^{2}}{\chi }.  \label{normom}
\end{equation}%
The energy defined by (\ref{emtn3}), (\ref{emtn5}) yields for a standing
wave (\ref{psiom}) the following expression 
\begin{equation}
\mathsf{E}=\frac{\chi ^{2}}{2m}\int \left[ \frac{1}{\mathrm{c}^{2}}\omega
^{2}\breve{\psi}\breve{\psi}^{\ast }+\kappa _{0}^{2}\breve{\psi}\breve{\psi}%
^{\ast }+\nabla \breve{\psi}\nabla \breve{\psi}^{\ast }+G_{a}\left( \breve{%
\psi}\breve{\psi}^{\ast }\right) \right] \,\mathrm{d}x.  \label{Estand}
\end{equation}%
The problem (\ref{eiga}), (\ref{normom}) has a sequence of solutions with
the corresponding sequence of frequencies $\omega $. Their energies $\mathsf{%
E}_{0\omega }$ are related to the frequency $\omega $ by the formula 
\begin{equation}
\mathsf{E}_{0\omega }=\chi \omega \left( 1+\Theta \left( \omega \right)
\right) ,\qquad \Theta \left( \omega \right) =\frac{a_{\mathrm{C}}^{2}}{%
2a^{2}}\frac{\omega _{0}^{2}}{\omega ^{2}},\qquad a_{\mathrm{C}}=\frac{\chi 
}{m\mathrm{c}},  \label{enomega}
\end{equation}%
where $a_{\mathrm{C}}=\frac{\lambda _{\mathrm{C}}}{2\pi }$ is the \emph{%
reduced Compton wavelength} of a particle with a mass $m$ and from now on we
assume that $\chi =\hbar $.

The above expression for the energy involves the size parameter $a$. It
coincides with the size of a \emph{free} electron in the absence of EM
fields. If our electron is placed in a strong EM field, its wave function $%
\psi $ significantly changes. In particular,\ in the hydrogen atom the
Coulomb field of a proton causes $\psi $ to become a perturbation of an
eigenstate of the linear hydrogen problem, \cite{BF6}. Therefore, its actual
size in the hydrogen atom is of order of the Bohr radius $a_{\mathrm{B}}$.
It is physically reasonable to assume that the Coulomb field of the proton
causes the distributed charge of the free electron to shrink, therefore $a$
should be larger than $a_{\mathrm{B}}$. \ An analysis shows that for
quantitative agreement with the classical hydrogen spectrum the value of $a$
for the electron should be at least $100a_{\mathrm{B}}$, \cite{BF6};
therefore, in (\ref{enomega}) $\ a_{\mathrm{C}}^{2}/a^{2}=\alpha ^{2}a_{%
\mathrm{B}}^{2}/a^{2}\lessapprox 10^{-8}$. This relatively large size of $a$
contrasts sharply with the concept of a point charge, but agrees well with
physical properties of electron both at atomic and macroscopic scales.\
Namely, for a larger value of $a$ the nonlinearity becomes smaller, and the
charge distribution at atomic scales becomes closer to the De Broglie wave, 
\cite{BF4}. Note also that the electrostatic potential generated by the free
charge with the size $a$ is very close to the exact Coulomb potential at a
macroscopic distance $R$ from the center of the charge, namely the
difference is of order $e^{-R^{2}/a^{2}}$ and is extremely small if $R$ is
greater than $a$.

\emph{\ }The Gaussian wave function 
\begin{equation}
\psi \left( t,\mathbf{x}\right) =\mathrm{e}^{-\mathrm{i}\omega
_{0}t}a^{-3/2}\pi ^{-3/4}\mathrm{e}^{-\left\vert x\right\vert ^{2}/2a^{2}}
\label{psioml}
\end{equation}%
with $\omega =\omega _{0}$ is the \emph{ground state} to the problem (it is
referred to as gausson in \cite{Bialynicki}). This state has the minimal
energy among all functions satisfying (\ref{normom}), hence it is stable. In
(\ref{enomega}) $a\gg a_{\mathrm{C}}$, and $\ \omega \geq \omega _{0}$ since 
$\omega _{0}$ corresponds to the ground state. Therefore, \emph{the relation
(\ref{enomega}) differs only slightly from the Planck-Einstein
energy-frequency relation} $\mathsf{E}=\hbar \omega $. Note that in the
non-relativistic version of our theory in (\cite{BF6}) the relation $\Delta 
\mathsf{E}=\hbar \Delta \omega $ is an exact identity for hydrogenic atoms.

By a change of variables the original nonlinear eigenvalue problem (\ref%
{eiga}) can be reduced to\ the following nonlinear eigenvalue problem with a
logarithmic nonlinearity with only one eigenvalue parameter $\xi $ and a
parameter-independent constraint: 
\begin{equation}
\nabla ^{2}\breve{\psi}_{1}=G_{1}^{\prime }\left( \left\vert \breve{\psi}%
\right\vert ^{2}\right) \breve{\psi}_{1}-\xi \breve{\psi}_{1},\;\int
\left\vert \breve{\psi}_{1}\right\vert ^{2}\,\mathrm{d}x=1.  \label{eipro1}
\end{equation}%
The parameter $\xi $ is related to the parameters in (\ref{eiga}) by the
formula 
\begin{equation}
\xi =\frac{a^{2}}{a_{\mathrm{C}}^{2}}\left( \frac{\omega ^{2}}{\omega
_{0}^{2}}-1\right) -\frac{1}{2}\ln \frac{\omega ^{2}}{\omega _{0}^{2}}.
\label{ksi}
\end{equation}%
The eigenvalue problem (\ref{eipro1}) has infinitely many solutions $\left(
\xi _{n},\psi _{1n}\right) $, $n=0,1,2...$, representing localized charge
distributions. The energy of $\psi _{n}$, $n>0$, is higher than the energy
of the Gaussian ground state which corresponds to $\xi =\xi _{0}=0$ and has
the lowest possible energy. These solutions coincide with critical points of
the energy\ functional under the constraint, for mathematical details see 
\cite{Cazenave83}, \cite{BerestyckiLions83I}, \cite{BerestyckiLions83II}.
The next two values of $\xi $ for the radial rest states are approximately $%
2.17$ and $3.41$ according to \cite{Bialynicki1}. Putting in (\ref{ksi}) the
values $\xi =\xi _{n}$ we can find the corresponding values of$\ \frac{%
\omega _{0}}{\omega }$, yielding for $a^{2}\gg a_{\mathrm{C}}^{2}$ the
following approximate formula\ 
\begin{equation}
\frac{\omega _{0}}{\omega }\simeq 1-\xi \frac{a_{\mathrm{C}}^{2}}{2a^{2}}.
\label{omksi1}
\end{equation}%
The difference of the energy of the higher states and the ground state
energy is small, it is of order $\hbar \omega _{0}\xi \frac{a_{\mathrm{C}%
}^{2}}{2a^{2}}=\xi \frac{a_{\mathrm{B}}^{2}}{a^{2}}\hbar \mathrm{c}R_{\infty
}$ where $\hbar \mathrm{c}R_{\infty }=m\mathrm{c}^{2}\frac{\alpha ^{2}}{2}$
is the Rydberg energy. Since we assume that $a_{\mathrm{B}}/a\lessapprox
10^{-4}$ the difference is comparable with the magnitude of the fine
structure in the hydrogen atom which is of order $m\mathrm{c}^{2}\alpha ^{4}$%
. This comparison shows that the cohesive forces generated by the
nonlinearity are relatively small.

Note that using the Lorentz invariance of the system one can easily obtain a
solution which represents the charge-field moving with a constant velocity $%
\mathbf{v}$ simply by applying to the rest solution $\left( \psi ,\varphi ,%
\mathbf{0}\right) $ the Lorentz transformation (see \cite{BF4}, \cite{BF5}
and the following Section \ref{Suniform}).

\begin{remark}
\label{Renergymass}The rest states of higher energies corresponding $\xi
=\xi _{n}$ with $n>0$ are unstable. Since the charges are coupled via
electromagnetic interactions, there is a possibility of the energy transfer
from charges with higher energies to the EM field, making such states very
improbable in normal circumstances. It is conceivable though that in a
system consisting of very many strongly interacting charges, such as, for
instance, an astronomical object, a significant quantity of such states may
be present contributing to the total energy and mass of such a system.
\end{remark}

\subsection{Uniform motion of a charge\label{Suniform}}

Consider now a free motion of a charge governed by the KG equation (\ref%
{KGex}) where the external fields vanish, that is $\varphi _{\mathrm{ex}}=0$,%
$\ \mathbf{A}_{\mathrm{ex}}=0$. Since the KG equation is relativistic
invariant, the solution can be obtained from a rest solution defined by (\ref%
{psiom}), (\ref{eiga}) by applying Lorentz transformation as in \cite{BF4}, 
\cite{BF6}. Hence, the solution to the KG equation (\ref{KGex}) which
represents a free particle that moves with velocity $\mathbf{v}$ is given by
the formula 
\begin{equation}
\psi \left( t,\mathbf{x}\right) =\psi _{\mathrm{free}}\left( t,\mathbf{x}%
\right) =\mathrm{e}^{-\mathrm{i}\left( \gamma \omega t-\mathbf{k}\cdot 
\mathbf{x}\right) }\breve{\psi}\left( \mathbf{x}^{\prime }\right) ,
\label{mvch1}
\end{equation}%
with $\breve{\psi}\left( \mathbf{x}^{\prime }\right) $ satisfying the
equation (\ref{eiga}), and%
\begin{equation}
\breve{\psi}\left( \mathbf{x}^{\prime }\right) =\breve{\psi}_{a}\left( 
\mathbf{x}^{\prime }\right) =a^{-3/2}\breve{\psi}_{1}\left( \mathbf{x}%
^{\prime }/a\right) ,  \label{psia}
\end{equation}%
\begin{equation}
\mathbf{x}^{\prime }=\mathbf{x}+\frac{\left( \gamma -1\right) }{v^{2}}\left( 
\mathbf{v}\cdot \mathbf{x}\right) \mathbf{v}-\gamma \mathbf{v}t,\ \quad 
\mathbf{k}=\gamma \omega \frac{\mathbf{v}}{\mathrm{c}^{2}},  \label{mvch4}
\end{equation}%
where $\gamma $ is the Lorentz factor 
\begin{equation}
\gamma =\left( 1-\mathbf{\beta }^{2}\right) ^{-1/2},\;\mathbf{\beta }=\frac{1%
}{\mathrm{c}}\mathbf{v}.  \label{gam}
\end{equation}%
All characteristics of a free charge can be explicitly written. Namely, the
charge density $\rho $ defined by the relation (\ref{paf6}) and the total
charge $\mathsf{E}$ equal respectively 
\begin{equation}
\rho =\gamma q\left\vert \breve{\psi}\left( \mathbf{x}^{\prime }\right)
\right\vert ^{2},\quad \bar{\rho}=\int \rho \left( \mathbf{x}\right) \,%
\mathrm{d}\mathbf{x}=q,  \label{rhofree}
\end{equation}%
\begin{equation}
\mathsf{E}=\gamma m\mathrm{c}^{2}\left( 1+\Theta \left( \omega \right)
\right) ,  \label{Eptfree}
\end{equation}%
where $\Theta \left( \omega \right) $ is given by (\ref{enomega}). The
current density$\mathbf{J}$, the total momentum $\mathbf{P}$ and the total
current $\mathbf{\bar{J}}$ for the free charge equal respectively%
\begin{equation}
\mathbf{J}=\frac{q}{m}\hbar \func{Im}\frac{\nabla \psi }{\psi }\left\vert
\psi \right\vert ^{2}=\gamma \hbar \frac{q}{m}\omega \frac{1}{\mathrm{c}^{2}}%
\mathbf{v}\left\vert \breve{\psi}\right\vert ^{2}\left( \mathbf{x}^{\prime
}\right) ,  \label{Pnr}
\end{equation}%
\begin{equation}
\mathbf{P}=\gamma m\mathbf{v}\left( 1+\Theta \left( \omega \right) \right)
,\quad \mathbf{\bar{J}}=\int \mathbf{J}\left( \mathbf{x}\right) \,\mathrm{d}%
\mathbf{x}=q\mathbf{v.}  \label{momfree}
\end{equation}%
The 4-vector $\left( \mathsf{E},\mathbf{P}\right) $ is a relativistic
energy-momentum 4-vector with the Lorentz invariant $\mathsf{E}^{2}-\mathrm{c%
}^{2}\mathbf{P}^{2}=\left( 1+\Theta \right) ^{2}m^{2}\mathrm{c}^{4}$. Hence,
based on the commonly used argument, \cite[Sec. 3.3]{Moller}, \cite[Sec. 37]%
{Pauli RT}, it is natural to define the \emph{rest mass} of the charge in
terms of the mass parameter $m$ by the formula 
\begin{equation}
m_{0}=m\left( 1+\Theta \left( \omega \right) \right) .  \label{Mfree}
\end{equation}%
A direct comparison shows that the above definition based on the Lorentz
invariance of uniformly moving free charge is fully consistent with the
definition of the inertial mass which was derived from the analysis of
accelerated motion of localized charges in external EM field in the previous
subsection, see Remark \ref{Rfreemass} for a more detailed discussion.

\subsection{The spectroscopic and inertial masses\label{SSpecmass}}

Our theory produces a description of the hydrogen atom, \cite{BF6}, \cite%
{BF7}, yielding in the non-relativistic case an asymptotic formula $%
E_{n}=-\hbar \mathrm{c}R_{\infty }/n^{2}$ for the hydrogen energy levels
with the factor $R_{\infty }$ given by the formula 
\begin{equation}
R_{\infty }=\frac{q^{4}m}{2\hbar ^{3}\mathrm{c}}.  \label{mrqa1}
\end{equation}%
In the relativistic case a more complex formula can be derived which
involves an equivalent of the Sommerfeld fine structure with the same factor 
$\hbar \mathrm{c}R_{\infty }$. The constant $R_{\infty }$ in (\ref{mrqa1})
coincides with\ the expression for the Rydberg constant if $m=m_{\mathrm{e}}$%
\ is the electron mass and $q$\ equals the electron charge. It seems,
therefore, natural to refer to the mass parameter $m$ as the \emph{%
spectroscopic mass}. We make a distinction between the spectroscopic mass
and\ the inertial mass since in our theory the mass parameter $m$ of a
charge is somewhat smaller than the inertial mass $m_{0}$ defined by the
formula (\ref{Mfree}) with $\omega =\omega _{0}$, namely 
\begin{equation}
m_{0}=m\left( 1+\frac{1}{2}\frac{a_{\mathrm{C}}^{2}}{a^{2}}\right) ,\text{ }%
a_{\mathrm{C}}=\frac{\hbar }{m\mathrm{c}}.  \label{mmaa1}
\end{equation}%
The difference $m_{0}-m$ depends evidently on the size parameter $a$. The
question stands now: is there any experimental evidence which shows that the
inertial mass $m_{0}$ and the spectroscopic mass $m$ may be different as in
our theory?

The quantum mechanics and the quantum electrodynamics allow to interpret the
spectroscopic data and extract from it the mass of electron known as the 
\emph{recommended value of the electron mass} $m=m_{\mathrm{e}}=A_{\mathrm{r}%
}\left( e\right) $. The recommended value of the electron mass $A_{\mathrm{r}%
}\left( e\right) $ when expressed in units $u$, can be found in \cite[Table
XLIX, p. 710]{Mohr6}:%
\begin{equation}
A_{\mathrm{r}}\left( e\right) =5.4857990943\left( 23\right) \times 10^{-4}%
\text{ }\left[ 4.2\times 10^{-10}\right] \text{ (recommended value),}
\label{masp2}
\end{equation}%
where, we remind the common convention, $5.4857990943$ represents the mean
value of the experimental data, $0.0000000023$ represents its standard
uncertainty (deviation) and $4.2\times 10^{-10}$ represents its relative
uncertainty. The value $A_{\mathrm{r}}\left( e\right) $ is found based on
very extensive spectroscopic data.

But there is also another class of measurements, namely the Penning\ trap
measurements, which can be considered as the most direct measurement of the
electron mass as the inertial one, and it gives the following mass value, 
\cite{Farnham}:%
\begin{equation}
A_{\mathrm{r}}\left( e\right) =5.485799111\left( 12\right) \times 10^{-4}%
\text{ }\left[ 2.1\times 10^{-9}\right] \text{ (Penning trap).}
\label{masp1}
\end{equation}%
Observe now that the value of the electron inertial mass coming from the
Penning trap measurement (\ref{masp1}) is larger than the spectroscopic mass
(\ref{masp2}), and the difference is statistically significant. Indeed,
using standard statistical analysis we obtain for the difference $\delta A_{%
\mathrm{r}}\left( e\right) =m_{0}-m$ the following%
\begin{equation}
m_{0}-m=\delta A_{\mathrm{r}}\left( e\right) =0.17(12)\times 10^{-11},\text{ 
}\left[ 3.1\times 10^{-9}\right] .  \label{masp1a}
\end{equation}%
If we take the\emph{\ recommended value} $A_{\mathrm{r}}\left( e\right) $
from (\ref{masp2}) as our "golden standard" then the mean value of the
Penning trap measurements from (\ref{masp1}) is more than 7 standard
deviations above the mean recommended value $A_{\mathrm{r}}\left( e\right)
=5.4857990943$, and the same is true for approximately one half of all
Penning trap measurements which are greater than the mean value. \emph{This
simple statistical calculation makes our point, namely that the inertial
mass }$m_{0}$\emph{\ as represented by Penning\ trap measurements (\ref%
{masp1}) is larger than the recommend value }$m$\emph{\ from (\ref{masp2}).}

We have not succeeded in finding in the existing literature an explanation
of the above mass difference, but in our theory such a difference including
its positive sign can be easily explained by the finite value of the
electron size parameter $a$. The resulting relation between the two masses
is given by the formula (\ref{mmaa1}) which can be recast as%
\begin{equation}
\frac{a_{\mathrm{C}}^{2}}{a^{2}}=2\frac{m_{0}-m}{m}.  \label{mmaa2}
\end{equation}%
From relations (\ref{masp1a}) and (\ref{masp2}) \ we obtain approximate
inequalities 
\begin{equation}
0.87\times 10^{-9}\preceq \frac{m_{0}-m}{m}\preceq 0.53\times 10^{-8}.
\label{mmaa3}
\end{equation}%
If we assign the effect to the non-zero value of $a$, using the formula (\ref%
{mmaa2}) we obtain%
\begin{equation}
0.97\times 10^{4}\preceq \frac{a}{a_{\mathrm{C}}}\preceq 2.4\times 10^{4}.
\label{mmaa4}
\end{equation}%
These inequalities are consistent with our prior assessments for $a$ to be
at least of order $10^{2}\alpha ^{-1}a_{\mathrm{C}}$. The mentioned
assessments are made in \cite{BF5}, \cite{BF6} based on the analysis of the
frequency spectrum of the hydrogen atom, see also a discussion after
relation (\ref{enomega}). Note that in the derivation of (\ref{mmaa4}) we
used an asymptotic formula for the spectrum from our theory of the hydrogen
atom. Consequently, the terms which were neglected may result in a much
larger interval\ for $a/a_{\mathrm{C}}$, and, hence, the relations (\ref%
{mmaa4}) are more for illustration purposes rather than for an actual
estimate of the ratio $a/a_{\mathrm{C}}$.

\subsection{Charge localization assumptions\label{secCL}}

The ergocenter obeys the relativistic version of Newton's equation if the
particle discrepancy terms $\mathbf{\delta }_{f},\mathbf{\delta }_{F},\delta
_{F}^{0}$ in the ergocenter equations (\ref{dtedelf}), (\ref{dterdel1}) can
be neglected. These terms would vanish exactly if the charge and
corresponding currents are localized exactly at the center $\mathbf{r}$, or
if $\mathbf{r}$ is a center of symmetry and the external EM potentials are
constant as in the case of a uniformly moving charge. In the general case we
may only expect that these terms vanish asymptotically in a certain limit,
namely, 
\begin{equation}
\mathbf{\delta }_{f}\rightarrow 0,\quad \mathbf{\delta }_{F}\rightarrow
0,\quad \delta _{F}^{0}\rightarrow 0.  \label{deldel0}
\end{equation}%
There are two kinds of quantities which enter the discrepancy terms. One
kind involves differences $\mathbf{E}_{\mathrm{ex}}\left( t,\mathbf{x}%
\right) -\mathbf{E}_{\mathrm{ex}}\left( t,\mathbf{r}\right) $ and $\mathbf{B}%
_{\mathrm{ex}}\left( t,\mathbf{x}\right) -\mathbf{B}_{\mathrm{ex}}\left( t,%
\mathbf{r}\right) $; they vanish if the fields are constant and are small if
the fields are almost constant. The magnitude of inhomogeneity of the EM
fields can be described by the typical length $R_{f}$ at which they vary
significantly. Hence, to ensure that the fields are almost constant near the
charge, we assume that $R_{f}$ is much larger than the charge size $a$ and
impose on the external field the following \emph{asymptotic local
homogeneity }condition: 
\begin{equation}
a/R_{f}\rightarrow 0.  \label{aac0}
\end{equation}%
Another kind of quantities which enter the discrepancy terms involve factors 
$\left( \mathbf{x}-\mathbf{r}\right) $ as in $\mathbf{\delta }_{F}$ in (\ref%
{delFdef}): 
\begin{gather*}
\int \frac{1}{\mathrm{c}}\left( \partial _{t}\left( \left( \mathbf{x}-%
\mathbf{r}\right) \rho \right) +\sum_{l}\partial _{l}\cdot \left( \left( 
\mathbf{x}-\mathbf{r}\right) \mathbf{J}^{l}\right) \right) \times \mathbf{B}%
_{\mathrm{ex}}\mathrm{d}\mathbf{x} \\
=\frac{1}{\mathrm{c}}\partial _{t}\int \rho \left( \mathbf{x}-\mathbf{r}%
\right) \times \mathbf{B}_{\mathrm{ex}}\mathrm{d}\mathbf{x-}\frac{1}{\mathrm{%
c}}\int \rho \mathbf{\left( \mathbf{x}-\mathbf{r}\right) }\times \partial
_{t}\mathbf{\mathbf{B}}_{\mathrm{ex}}d\mathbf{\mathbf{x}}+\int \frac{1}{%
\mathrm{c}}\int \sum_{l}\mathbf{J}^{l}\left( \mathbf{x}-\mathbf{r}\right)
\times \partial _{l}\mathbf{B}_{\mathrm{ex}}\mathrm{d}\mathbf{x.}
\end{gather*}%
These quantities vanish for spatially constant $\mathbf{\mathbf{B}}_{\mathrm{%
ex}},\mathbf{E}_{\mathrm{ex}}$ if $\rho $ and every component of $\mathbf{J}$
is center-symmetric with respect to $\mathbf{r}$. Similar quantities are
present in$\ \delta _{F}^{0}$ and $\mathbf{\delta }_{f}$ defined by (\ref%
{delf0}), (\ref{emtn10}). For example 
\begin{equation}
\delta _{f}^{i}=\int \left( x^{i}-r^{i}\right) \left( \rho \mathbf{E}_{%
\mathrm{ex}}+\frac{1}{\mathrm{c}}\mathbf{J}\times \mathbf{B}_{\mathrm{ex}%
}\right) ^{i}\mathrm{d}\mathbf{x}
\end{equation}%
vanish under the same central symmetry assumption. Hence, to satisfy the
charge localization condition (\ref{deldel0}), it is sufficient to impose
two separate requirements: (i) the asymptotic condition (\ref{aac0}) and
(ii) asymptotic central symmetry of $\rho $ and components of $\mathbf{J}$.

To clarify the meaning of the above localization conditions, let us look at
the simplest case where $\rho $ and $\mathbf{J}$ are derived for the uniform
motion of a free particle described by the solution $\psi \left( t,\mathbf{x}%
\right) =\psi _{\mathrm{free}}\left( t,\mathbf{x}\right) $ considered in
Section \ref{Suniform}. The solution has the following properties:

(i) the energy density $u$\ is center-symmetric with respect to $\mathbf{r}%
\left( t\right) =\mathbf{v}t$, hence the ergocenter coincides with $\mathbf{r%
}\left( t\right) $;

(ii) the charge density $\rho $ is given by (\ref{rhofree}), and according
to (\ref{psia}) it converges to $q\delta \left( \mathbf{x}-\mathbf{r}\right)
\ $as $a\rightarrow 0$ where $\delta \left( \mathbf{x}\ \right) $ is the
Dirac delta-function;

(iii) the current $\mathbf{J}$ is given by (\ref{Pnr}), its components are
center-symmetric and converge to the corresponding components of $q\mathbf{v}%
\delta \left( \mathbf{x}-\mathbf{r}\right) $.

Hence, the localization assumptions (\ref{deldel0}) are fulfilled for $\rho $
and $\mathbf{P}$ derived for $\psi _{\mathrm{free}}\left( t,\mathbf{x}%
\right) \ $for general fields $\mathbf{E}_{\mathrm{ex}},\mathbf{B}_{\mathrm{%
ex}}$ which are regular near $\mathbf{r}$ when $a/R_{f}\rightarrow 0$. If
the motion is almost uniform, namely if the external fields are not too
strong, and the solution $\psi \left( t,\mathbf{x}\right) $ of (\ref{KGex})
is close to $\psi _{\mathrm{free}}\left( t,\mathbf{x}\right) $, we may
expect that such a solution also satisfies the localization conditions, we
present an example in Section \ref{secAR}.

Now let us briefly discuss condition (\ref{aac0}). Suppose that\ the
corresponding forces, according to (\ref{rederg2}), are of the order $\hat{f}%
_{\mathrm{Lor}}$ where $\hat{f}_{\mathrm{Lor}}$ is a typical magnitude of
the Lorentz force $\mathbf{f}_{\mathrm{Lor}}\left( t,\mathbf{r}\right) $.
The spatial scale $R_{f}$ at which the forces associated with the
electromagnetic fields $\mathbf{E}_{\mathrm{ex}},\mathbf{B}_{\mathrm{ex}}\ $%
vary by the same order of magnitude as $\hat{f}_{\mathrm{Lor}}$ can be
defined as follows: 
\begin{equation}
\frac{1}{R_{f}}=\frac{q}{\hat{f}_{\mathrm{Lor}}}\max_{\left\vert
x-r\right\vert \leq \theta a}\left( \frac{1}{\mathrm{c}}\left\vert \nabla 
\mathbf{B}_{\mathrm{ex}}\right\vert +\left\vert \nabla \mathbf{E}_{\mathrm{ex%
}}\right\vert \right)  \label{onerf}
\end{equation}%
with $\theta \gg 1$. In view of this definition, condition (\ref{aac0})\
ensures that the variability of EM fields causes a vanishing perturbation to
the Lorentz force in (\ref{rederg2}).

The estimates of smallness of the discrepancy terms can be made in certain
asymptotic regimes, but they are laborious, and we would like to make some
guiding comments. The KG equation (\ref{KG1D}) evidently involves five
parameters, namely, $\mathrm{c}$,$\hbar ,m$,\ $q,a$, and\ the external
fields also can depend on parameters. The limits (\ref{deldel0}) can be
considered simultaneously with certain combinations of the parameters
tending to their limits, and the limits in (\ref{deldel0}) have to be taken
together with these parameter limits. Note that all the quantities which
enter equations (\ref{dterdel1}) (\ref{dtedelf}) are the integrals of
certain densities over the entire space. The integrals over the entire space
obviously are the limits of the integrals over the domains $\left\vert 
\mathbf{x}-\mathbf{r}\right\vert <\theta a$ with $\theta \rightarrow \infty $%
, and the value of $\theta $ in the asymptotic regimes can be related with
the values of the other parameters mentioned. An example of such asymptotic
situation with nontrivial relations between the parameters is given in the
following section.

\section{Relativistic accelerated motion of a\ wave-corpuscle \label{secAR}}

Up to now, we were considering relativistic features of an accelerating
charge assuming its localization. The localization assumptions, though
natural, are rather technical when it comes to rigorous treatment. With that
in mind, we study in this section a particular case where the dynamical
problem is non-trivial and produces a wide variety of accelerated
relativistic motions of the charge which are simple enough for a detailed
analysis and the verification of the localization assumptions. The analysis
is still rather involved, and that comes at no surprise since for general
external EM fields the KG equation (\ref{KGex}) has no closed form
solutions. We succeeded though in finding a large family of non-trivial
regimes for which we obtain almost explicit representations of solutions
allowing for a detailed study of the relativistic features of the charge
accelerating in an external EM field.

\subsection{Rectilinear charge motion}

In the previous section we studied dynamics of localized accelerating waves
in general EM fields. The relativistic point dynamics was derived under the
assumption that the localization conditions (\ref{deldel0}) hold. Here we
present an example of an accelerating charge when its localization can be
maintained in the strongest possible form. Namely, we consider a regime when
the accelerating charge \emph{exactly preserves its Gaussian shape} up to
the Lorentz contraction; we call such a solution of the KG equation a
wave--corpuscle. The charge wave function is similar to $\psi _{\mathrm{free}%
}\left( t,\mathbf{x}\right) $ in (\ref{mvch1})\ and consequently it is
localized about $\mathbf{r}\left( t\right) $, but its velocity $\mathbf{v}%
=\partial _{t}\mathbf{r}$ is not constant and the charge has a non-zero
acceleration. \emph{The shape of the wave in an accelerated regime is
exactly the same as for a free particle, and it is only the phase factor
that is affected by the acceleration caused by the external force}.\emph{\ }%
Such an accelerated motion with a fixed shape is possible only for a
properly chosen potential $\varphi _{\mathrm{ex}}$. We consider here the
simplest but still non-trivial case where the charge moves and accelerates
in the direction of the axis $x_{3}$ with the potential $\varphi _{\mathrm{ex%
}}$ being a function of only the variable $x_{3}$ and the time $t$, and
there is no external magnetic field, that is $\mathbf{A}_{\mathrm{ex}}=0$.

When the external potential $\varphi _{\mathrm{ex}}$ depends only on $%
t,x_{3} $, the equation (\ref{KGex}) in the three dimensional space with a
logarithmic nonlinearity (\ref{paf3}) can be exactly reduced to a problem in
one dimensional space by the following substitution 
\begin{equation}
\psi =\pi ^{-1/2}a^{-1}\exp \left( -\frac{1}{2a^{2}}\left(
x_{1}^{2}+x_{2}^{2}\right) \right) \psi _{1D}\left( t,x_{3}\right) ,
\label{psi1d}
\end{equation}%
We obtain a reduced 1D KG equation\ for $\psi =\psi _{1D}\left(
t,x_{3}\right) $ with \emph{one spatial variable}: 
\begin{equation}
-\frac{1}{\mathrm{c}^{2}}\tilde{\partial}_{t}\tilde{\partial}_{t}\psi
+\partial _{3}^{2}\psi -G_{a1}^{\prime }\left( \psi ^{\ast }\psi \right)
\psi -\kappa _{0}^{2}\psi =0.  \label{KG1D}
\end{equation}%
where 
\begin{equation*}
\kappa _{0}=\frac{m\mathrm{c}}{\hbar },\quad \tilde{\partial}_{t}=\partial
_{t}+\frac{\mathrm{i}q}{\hbar }\varphi _{\mathrm{ex}},\quad \varphi _{%
\mathrm{ex}}=\varphi _{\mathrm{ex}}\left( t,x_{3}\right) ,
\end{equation*}%
and the $1D$ logarithmic nonlinearity has the form 
\begin{equation}
G_{a1}^{\prime }\left( \left\vert \psi \right\vert ^{2}\right) =-a^{-2}\left[
\ln \left( \pi ^{1/2}\left\vert \psi \right\vert ^{2}\right) +1\right]
-a^{-2}\ln a.  \label{logb}
\end{equation}%
This equation has a Gaussian as a rest solution, 
\begin{equation}
\breve{\psi}\left( x_{3}\right) =\pi ^{-1/4}a^{-1/2}\mathrm{e}%
^{-x_{3}^{2}/2a^{2}}.  \label{psi0}
\end{equation}%
From now on we write $x$ instead of $x_{3}$ for the notational simplicity.

We consider the center location $r\left( t\right) $ to be a given function.
We also assume that the motion is translational. \ Namely, the solution $%
\psi \left( t,x\right) $ to (\ref{KG1D}) \ has the following Gaussian form:

\begin{gather}
\psi =a^{-1/2}\mathrm{e}^{-\mathrm{i}S_{0}\left( t,y\right) }\mathring{\psi}%
(a^{-1}\gamma _{0}y),\quad y=x-r\left( t\right) ,  \label{psigauss} \\
\mathring{\psi}\left( t,z\right) =\pi ^{-1/4}\gamma ^{-1/2}\gamma _{0}^{1/2}%
\mathrm{e}^{-z^{2}/2},  \notag
\end{gather}%
where $\gamma =\gamma \left( t\right) \ $ is the Lorentz factor, $\gamma
_{0}=\gamma \left( 0\right) $, the phase \ $S_{0}\left( t,y\right) $ has to
be determined. \ We look for such a potential $\varphi _{\mathrm{ex}}$\ that
the wave $\psi $ defined by (\ref{psigauss}) is an exact solution of the KG\
equation (\ref{KG1D}). If such a potential is found, the function $\psi
\left( t,x\right) $ describes an accelerating charge with the strongest
possible localization, namely, a charge with a fixed shape $\left\vert \psi
\right\vert $ and arbitrarily small size $a$. \ 

\subsection{Mild acceleration regime}

Below we provide an example of a regime which allows simultaneously
localization and acceleration, with a general charge trajectory $r\left(
t\right) $ subjected to certain regularity conditions. The KG equation (\ref%
{KG1D}) involves five parameters, namely the speed of light $\mathrm{c}$,
Planck constant $\hbar $, the mass parameter $m,$ the charge value $q$ and
the charge size $a$. A combination of the parameters which is important in
our analysis is the reduced Compton wavelength $a_{_{\mathrm{C}}}=\frac{%
\hbar }{m\mathrm{c}}$. Now we describe relations between the parameters
which allow for localized accelerating charges.

We begin with introducing a class of admissible trajectories $r\left(
t\right) $. \ We assume that the charge does not undergo violent
accelerations and suppose a regular dependence of the normalized velocity $%
\beta $ on the dimensionless time $\tau =\mathrm{c}t/a$, namely 
\begin{equation}
\max_{\tau }\left( \left\vert \partial _{\tau }\beta \right\vert +\left\vert
\partial _{\tau }^{2}\beta \right\vert +\left\vert \partial _{\tau
}^{3}\beta \right\vert \right) \leq \epsilon ,\quad \text{where }\tau =\frac{%
\mathrm{c}}{a}t.  \label{betep0}
\end{equation}%
We also assume that the velocity $v=\partial _{t}r,$ as well as its
variation, \ \ is smaller than the speed of light $\mathrm{c}$, namely 
\begin{equation}
\frac{1}{\mathrm{c}}\max_{t}\left\vert v\left( t\right) \right\vert \leq
\epsilon _{1},\qquad \frac{1}{\mathrm{c}}\max_{t}\left\vert v\left( t\right)
-v\left( 0\right) \right\vert \leq \epsilon _{1}\qquad \epsilon _{1}<1.
\label{cv10}
\end{equation}%
Obviously, the variation of $\beta $ and of the Lorentz factor $\gamma $
remains bounded for all $\tau $, that is 
\begin{equation}
\left\vert \gamma _{0}-\gamma \left( \tau \right) \right\vert \leq C\epsilon
_{1},\qquad \left\vert \beta _{0}-\beta \left( \tau \right) \right\vert \leq
\epsilon _{1},  \label{eps1}
\end{equation}%
where $\gamma _{0}=\gamma \left( 0\right) ,$ $\beta _{0}=\beta \left(
0\right) $. \ Observe that the above assumptions are consistent with
significant changes of velocity $v$ in asymptotic regimes where $\mathrm{c}%
\rightarrow \infty ,$ $v\rightarrow \infty $, $v/\mathrm{c}\rightarrow 
\limfunc{const}$.\ Note that condition (\ref{eps1}) does not follow from (%
\ref{betep0}). It can can be seen from the following example of oscillatory
motion clarifying the above conditions.

Let $v\left( t\right) $ be defined by 
\begin{gather}
v\left( t\right) =\mathrm{c}\beta _{0}+\bar{v}_{1}\sin ^{2}\left( \frac{%
\mathrm{c}}{a}\eta t\right) =\mathrm{c}\beta _{0}+\bar{v}_{1}\sin ^{2}\left(
\eta \tau \right) \text{ for }t\geq 0,  \label{vosc} \\
v\left( t\right) =\mathrm{c}\beta _{0}\text{ for }t\leq 0,  \notag
\end{gather}%
where $\bar{v}_{1}$ is the amplitude of the variable part of velocity, $\eta 
$ is a dimensionless frequency, $\beta _{0}<1$ is the initial normalized
velocity. We readily obtain the following estimate of the normalized
velocity variation:%
\begin{equation*}
\frac{1}{\mathrm{c}}\left\vert v\left( t\right) -v\left( 0\right)
\right\vert \leq \frac{1}{\mathrm{c}}\bar{v}_{1},
\end{equation*}%
and to satisfy (\ref{cv10}) and (\ref{betep0}) \ we set 
\begin{equation}
\frac{1}{\mathrm{c}}\bar{v}_{1}=\epsilon _{1},\qquad \frac{1}{\mathrm{c}}%
\bar{v}_{1}\left( \eta +\eta ^{2}+\eta ^{3}\right) =\epsilon .  \label{vep1}
\end{equation}%
Since the parameter $\eta $ can depend on $\epsilon _{1}$, the boundedness
of $\epsilon _{1}$ and $\epsilon $ \ do not follow one from another. The
above conditions imply a bound on the acceleration, 
\begin{equation*}
\left\vert \partial _{t}v\right\vert =\eta \bar{v}_{1}\mathrm{c}%
a^{-1}\left\vert \sin \left( 2\eta \mathrm{c}t/a\right) \right\vert \leq
\epsilon \mathrm{c}^{2}/a,
\end{equation*}%
which involves a large factor $\mathrm{c}^{2}/a$; therefore, the
acceleration can be large even if the parameter $\epsilon \ $is small.

Let us consider now the following important \emph{dynamical problem}:
determine the external electric field which generates a purely translational
motion described by (\ref{psigauss}). The size parameter $a$ sets up a
natural microscopic length scale. We assume that the velocity $v\left(
t\right) $ satisfies conditions (\ref{betep0}) and (\ref{cv10}), and the
parameters of the problem satisfy the following two restrictions. First, we
assume that the Compton wavelength defined in (\ref{enomega}) is much
smaller than the size parameter $a$, namely%
\begin{equation}
\zeta =\frac{a_{\mathrm{C}}}{a}\ll 1.  \label{zetzet0}
\end{equation}%
The second condition relates $\zeta $ to the parameters $\epsilon $ and $%
\epsilon _{1}$ from (\ref{betep0}), (\ref{cv10}):%
\begin{equation}
\frac{\epsilon _{1}}{\epsilon }\zeta ^{2}\ll 1.  \label{epzetep}
\end{equation}%
For a given trajectory $r\left( t\right) $ we find then a\ potential $%
\varphi _{\mathrm{ex}}$ such that the Gaussian wave function with the center 
$r\left( t\right) $ is an \emph{exact} solution of (\ref{KG1D}) in the strip 
$\Xi \left( \theta \right) =\left\{ \left\vert x-r\left( t\right)
\right\vert \leq \theta a\right\} $\ around the trajectory, and $\theta $
grows to infinity as $\zeta \rightarrow 0$.\ The solution $\psi $ is similar
to the free solution but it allows for an acceleration. We will show that
the trajectory $r\left( t\right) $ relates to the potential according to the
relativistic version of Newton's law 
\begin{equation}
\partial _{t}\left( m\gamma v\right) +q\partial _{x}\varphi _{\mathrm{ac}%
}\left( t,r\right) =0.  \label{new11}
\end{equation}%
Here $\varphi _{\mathrm{ac}}\left( x\right) $ is the leading part of the
potential $\varphi _{\mathrm{ex}}$ which causes the acceleration of the
charge, we call it "accelerating" potential, it does not depend on the small
parameter $\zeta $. The remaining part of the external potential $\varphi _{%
\mathrm{ex}}$ in the KG equation (\ref{KG1D}) is a small "balancing"
potential 
\begin{equation*}
\varphi _{2}\left( t,x,\zeta \right) =\varphi _{\mathrm{ex}}\left( t,x;\
\zeta \right) -\varphi _{\mathrm{ac}}\left( t,x\right) ,
\end{equation*}%
which allows the charge to exactly preserve its form as it accelerates. The
balancing potential vanishes asymptotically, that is $\varphi _{2}\left(
t,x;\zeta \right) \rightarrow 0$ as $\zeta \rightarrow 0$, and the forces it
produces also become vanishingly small compared with the Lorentz force $%
q\partial _{x}\varphi _{\mathrm{ac}}\left( x\right) $\ in the strip $\Xi
\left( \theta \right) $. The potential $\varphi _{\mathrm{ex}}\left(
t,x\right) $ converges to its accelerating part $\varphi _{\mathrm{ac}}$ in
the strip $\Xi \left( \theta \right) $ with $\theta \rightarrow \infty $ \
as $\zeta \rightarrow 0$. In view of (\ref{new11}) it is natural to treat $%
\varphi _{2}$ as a perturbation which does not affect the acceleration, but
is responsible for the exact preservation of the shape of charge
distribution during its evolution. In the asymptotic limit $\zeta
\rightarrow 0$ we have in (\ref{new11}) 
\begin{equation}
\partial _{x}\varphi _{\mathrm{ex}}\left( r\left( t\right) \right)
\rightarrow \partial _{x}\varphi _{\mathrm{ac}}\left( t,r\left( t\right)
\right) .
\end{equation}%
The summary of what we intend to fulfill is as follows. For any given
trajectory $r\left( t\right) $ we construct a potential $\varphi _{\mathrm{ex%
}}$ which makes the Gaussian wave with the center $r\left( t\right) $ to be
an exact solution of the field equation\ in the widening strip $\Xi \left(
\theta \right) $, $\theta \rightarrow \infty $. Since the shape of $%
\left\vert \psi \right\vert $ is preserved, such a function is localized
around $r\left( t\right) $. We can show that in the considered regimes the
typical spatial scale $R$ of inhomogeneity of the constructed electric field
tends to infinity, $R/a\rightarrow \infty $, and hence (\ref{aac0}) is
fulfilled.

The implementation of the outlined above approach is provided below and here
is yet another look at it. We construct the external field which causes the
accelerated motion of a charge distribution with a fixed Gaussian shape; the
distribution exactly satisfies the relativistic covariant KG equation in a
wide strip about the wave center trajectory. Consequently, it provides an
example for the relativistic dynamics of a charge wave function with a fixed
shape. The possibility of a uniform global motion without acceleration is
well-known, see Section \ref{Suniform} and \cite{BenciF}. The fact that
relativistic acceleration imposes restrictions on the spatial extension of
rigid bodies was noted in a different setting in \cite{EriksenML82}.

\subsection{Equation in a moving frame}

As the first step of the analysis we rewrite the KG equation (\ref{KG1D}) in
a moving frame around $r$. We take $r\left( t\right) $ as the new origin and
make the following change of variables 
\begin{equation}
x_{3}=r\left( t\right) +y,\quad \psi \left( t,x\right) =\psi ^{\prime
}\left( t,y\right) ,\quad v=\partial _{t}r.  \label{xry}
\end{equation}%
The\ 1D KG equation (\ref{KG1D}) then takes the form 
\begin{gather}
-\frac{1}{\mathrm{c}^{2}}\left( \partial _{t}+\frac{\mathrm{i}q\varphi _{%
\mathrm{ex}}}{\hbar }-v\mathbf{\partial }_{y}\right) \left( \partial _{t}+%
\frac{\mathrm{i}q\varphi _{\mathrm{ex}}}{\hbar }-v\mathbf{\partial }%
_{y}\right) \psi ^{\prime }  \label{KGy} \\
+\partial _{3}^{2}\psi ^{\prime }-G_{a}^{\prime }\left( \psi ^{\prime \ast
}\psi ^{\prime }\right) \psi ^{\prime }-\frac{1}{a_{C}^{2}}\psi ^{\prime }=0,
\notag
\end{gather}%
where $v\left( t\right) $ is a given function of time. We write the external
potential $\varphi _{\mathrm{ex}}\ $which produces the motion in the form 
\begin{equation}
\varphi _{\mathrm{ex}}\left( t,x\right) =\varphi _{\mathrm{ac}}\left(
t,y\right) +\varphi _{2}\left( t,y,\zeta \right) ,  \label{fiex}
\end{equation}%
where the \emph{accelerating potential is linear in }$y$\ 
\begin{equation*}
\varphi _{\mathrm{ac}}=\varphi _{\mathrm{0}}\left( t\right) +\varphi _{%
\mathrm{ac}}^{\prime }y,
\end{equation*}%
and $\varphi _{2}\left( t,y\right) $ is a \emph{balancing potential}, which,
as we will show, is small. The 1D logarithmic nonlinearity $G_{a}^{\prime
}=G_{a1}^{\prime }$ is defined by (\ref{logb}). We assume that the solution $%
\psi \left( t,x\right) $ has the Gaussian form, namely 
\begin{equation}
\psi ^{\prime }=\mathrm{e}^{\mathrm{i}\frac{\omega _{0}}{\mathrm{c}^{2}}%
\gamma vy-\mathrm{i}s\left( t\right) -\mathrm{i}S\left( t,y\right) }\Psi
^{\prime },\quad \omega _{0}=\frac{\mathrm{c}}{a_{\mathrm{C}}},
\label{psipsi}
\end{equation}%
where we define $\Psi ^{\prime }$ by an explicit expression 
\begin{equation}
\Psi ^{\prime }=a^{-1/2}\mathring{\psi}_{1}\left( a^{-1}\gamma _{0}y\right) ,
\label{psigam}
\end{equation}%
where $\gamma _{0}$ is the Lorentz factor defined by (\ref{gam}) from (\ref%
{eps1}) and%
\begin{equation}
\mathring{\psi}_{1}\left( t,z\right) =\pi ^{-1/4}\mathrm{e}^{\sigma
-z^{2}/2},  \label{psi1sig}
\end{equation}%
\begin{equation}
\sigma =\sigma \left( t\right) =\frac{1}{2}\ln \frac{\gamma _{0}}{\gamma }.
\label{siggam}
\end{equation}

Using (\ref{eps1}) and (\ref{betep0}), we observe that $\sigma $ is bounded,
namely 
\begin{equation}
\sigma =O\left( \epsilon _{1}\right) ,\quad \partial _{\tau }\sigma =O\left(
\epsilon \right) ,\quad \partial _{\tau }^{2}\sigma =O\left( \epsilon
\right) .  \label{sigeps1}
\end{equation}%
Substitution of (\ref{psipsi}) into (\ref{KGy}) produces 
\begin{gather}
-\frac{1}{\mathrm{c}^{2}}\left( \partial _{t}+\mathrm{i}\partial _{t}\left(
\gamma v\right) \frac{\omega _{0}}{\mathrm{c}^{2}}y-\mathrm{i}\partial _{t}s+%
\frac{\mathrm{i}q\varphi _{\mathrm{0}}}{\hbar }-\mathrm{i}v^{2}\frac{\omega
_{0}}{\mathrm{c}^{2}}\gamma \right.  \notag \\
\left. -\mathrm{i}\partial _{t}S+\mathrm{i}v\partial _{y}S+\frac{\mathrm{i}%
q\varphi _{\mathrm{ac}}^{\prime }}{\hbar }y+\frac{\mathrm{i}q\varphi _{2}}{%
\hbar }-v\partial _{y}\right) ^{2}\Psi ^{\prime }  \notag \\
+\left( \partial _{y}-\mathrm{i}\partial _{y}S+\mathrm{i}v\frac{\omega _{0}}{%
\mathrm{c}^{2}}\gamma \right) ^{2}\Psi ^{\prime }-G_{a}^{\prime }\left(
\left\vert \Psi ^{\prime }\right\vert ^{2}\right) \Psi ^{\prime }-\kappa
_{0}^{2}\Psi ^{\prime }=0.  \label{KGy1}
\end{gather}%
We show below that $\varphi _{2}$ and $S$ are of order $\zeta ^{2}\left(
\epsilon _{1}+\epsilon \right) $. To eliminate the leading (independent of
the small parameter $\zeta $)\ terms in the above equation, we require that
the following two equations are fulfilled: 
\begin{equation}
-\partial _{t}s+\frac{q\varphi _{\mathrm{0}}}{\hbar }-v^{2}\frac{\omega _{0}%
}{\mathrm{c}^{2}}\gamma =-\gamma \omega _{0},  \label{sprime}
\end{equation}%
and 
\begin{equation}
m\partial _{t}\left( \gamma v\right) +q\varphi _{\mathrm{ac}}^{\prime }=0.
\label{Newt}
\end{equation}%
Obviously, the expressions in (\ref{KGy1}) eliminated via equations (\ref%
{sprime}) and (\ref{Newt}) do not depend on $\zeta $.\ The equation (\ref%
{Newt}) determines the essential part $\varphi _{\mathrm{ac}}^{\prime }y$ of
the accelerating potential; the constant part $\varphi _{\mathrm{0}}\left(
t\right) $ of the accelerating potential can be prescribed arbitrarily,
since we always can choose the phase shift $s\left( t\right) $ so that (\ref%
{sprime}) is fulfilled. Equation (\ref{Newt}) evidently coincides with the 
\emph{relativistic law of motion} (\ref{new11}).\ 

The above conditions annihilate in the equation (\ref{KGy1}) all the terms
which do not vanish as $\zeta $ tends to zero; we show this in the following
two sections. As a first step, we simplify equation (\ref{KGy}) using
equations (\ref{sprime}) and (\ref{Newt}), and obtain the following
equation: 
\begin{gather}
-\frac{1}{\mathrm{c}^{2}}\left( \partial _{t}-\mathrm{i}\gamma \frac{\mathrm{%
c}}{a_{\mathrm{C}}}-\mathrm{i}\partial _{t}S+\mathrm{i}v\partial _{y}S+\frac{%
\mathrm{i}q\varphi _{2}}{\hbar }-v\partial _{y}\right) ^{2}\Psi ^{\prime } 
\notag \\
+\left( \partial _{y}-\mathrm{i}\partial _{y}S+\mathrm{i}\frac{1}{a_{\mathrm{%
C}}}\frac{v}{\mathrm{c}}\gamma \right) ^{2}\Psi ^{\prime }-G_{a}^{\prime
}\left( \left\vert \Psi ^{\prime }\right\vert ^{2}\right) \Psi ^{\prime }-%
\frac{1}{a_{\mathrm{C}}^{2}}\Psi ^{\prime }=0.  \label{Kpsy}
\end{gather}%
In the following sections \ we find small $\varphi _{2}$ and $S$ which are
of order $\zeta ^{2}\left( \epsilon _{1}+\epsilon \right) $.

\subsection{Equations for auxiliary phases}

In this subsection we introduce two auxiliary phases, and then reduce the
problem of determination of the potential to a simpler first-order partial
differential equation for one unknown phase. Solving such an equation can be
further reduced to integration along characteristics and allows a rather
detailed mathematical analysis. Here we restrict ourselves to the simplest
steps in the analysis, but the introduced construction can be used for a
much more detailed analysis of the relativistic interaction of the EM field
with a rigid charge.

It is convenient to introduce rescaled dimensionless variables $z,\tau $:%
\begin{gather}
\tau =\frac{\mathrm{c}}{a}t,\quad z=\frac{\zeta }{a_{\mathrm{C}}}y=\frac{1}{a%
}y,\quad \Psi =a^{1/2}\Psi ^{\prime },  \label{zztau} \\
\partial _{y}=\frac{1}{a}\partial _{z}=\frac{\zeta }{a_{\mathrm{C}}}\partial
_{z},\quad \partial _{t}=\frac{\mathrm{c}}{a}\partial _{\tau }=\zeta \frac{%
\mathrm{c}}{a_{\mathrm{C}}}\partial _{\tau }.  \notag
\end{gather}%
We introduce auxiliary phases $Z$ and $\Phi $, 
\begin{gather}
Z=\zeta \mathbf{\partial }_{z}S,  \label{denZ} \\
\Phi =-\zeta \mathbf{\partial }_{\tau }S+\zeta \beta \mathbf{\partial }_{z}S+%
\frac{qa_{\mathrm{C}}\varphi _{2}}{\mathrm{c}\hbar };  \label{denZ2}
\end{gather}%
they will be our new unknown variables. These auxiliary phases determine the
balancing potential $\varphi _{2}$ which we intend to make vanishingly
small. Obviously, if we find $Z$ and $\Phi $, we can find $S$ by the
integration in $z$ and we set $S=0$ at $z=0$. After that $\varphi _{2}$ can
be found from (\ref{denZ2}). Consequently, to find a small $\varphi _{2}$ we
need to find small $Z$ \ and $\Phi $.

Equation (\ref{Kpsy}) takes the following form:%
\begin{gather}
-\left( \zeta \partial _{\tau }+\mathrm{i}\Phi -\mathrm{i}\gamma -\beta
\zeta \partial _{z}\right) ^{2}\Psi  \label{eqips} \\
+\left( \zeta \partial _{z}-\mathrm{i}Z+\mathrm{i}\beta \gamma \right)
^{2}\Psi -\zeta ^{2}G_{1}^{\prime }\left( \Psi ^{\ast }\Psi \right) \Psi
-\Psi =0.  \notag
\end{gather}%
We look for a solution of (\ref{eqips}) in the strip $\Xi $ in time-space: 
\begin{equation}
\Xi =\left\{ \left( \tau ,z\right) :-\infty \leq t\leq \infty ,\quad
\left\vert z\right\vert \leq \theta \right\}  \label{Ksi}
\end{equation}%
where $\theta $ is a large number. Note that $\left\vert \Psi \right\vert
=\pi ^{-1/4}\mathrm{e}^{-\gamma _{0}^{2}z^{2}/2}$ is smaller than $\pi
^{-1/4}\mathrm{e}^{-\theta ^{2}/2}$\ outside $\Xi $ and is extremely small
for large $\theta $, and later we are going to make $\theta $ arbitrarily
large. We expand (\ref{eqips}) with respect to $\Phi ,Z$ yielding 
\begin{gather}
Q\Psi -\mathrm{i}\Phi \left( \zeta \partial _{\tau }-\mathrm{i}\gamma -\beta
\zeta \partial _{z}\right) \Psi -\mathrm{i}\left( \zeta \partial _{\tau }-%
\mathrm{i}\gamma -\beta \zeta \partial _{z}\right) \left( \Phi \Psi \right)
+\Phi ^{2}\Psi  \label{Qeq} \\
+\mathrm{i}Z\left( \zeta \partial _{z}+\mathrm{i}\beta \gamma \right) \Psi +%
\mathrm{i}\left( \zeta \partial _{z}+\mathrm{i}\beta \gamma \right) \left(
Z\Psi \right) -Z^{2}\Psi =0,  \notag
\end{gather}%
where we denote by $Q\Psi $ the term which does not involve $\Phi $ and $Z$
explicitly: 
\begin{equation}
Q=\frac{1}{\Psi }\left( -\left( \zeta \partial _{\tau }-\mathrm{i}\gamma
-\beta \zeta \partial _{z}\right) ^{2}\Psi +\left( \zeta \partial _{z}+%
\mathrm{i}\beta \gamma \right) ^{2}\Psi \right) -\zeta ^{2}G_{1}^{\prime
}\left( \Psi ^{\ast }\Psi \right) -1.  \label{Qdef}
\end{equation}%
\bigskip Since (\ref{siggam}) holds, the imaginary part of $Q$ is zero, 
\begin{equation}
\func{Im}Q=2\zeta \gamma \partial _{\tau }\sigma +\zeta \partial _{\tau
}\gamma =0.  \label{imips}
\end{equation}%
Hence $Q$\ is real and we have%
\begin{equation}
Q=\func{Re}Q=\left( -\frac{1}{\Psi }\partial _{\tau }^{2}\Psi +\partial
_{\tau }\beta \frac{1}{\Psi }\partial _{z}\Psi +\frac{2}{\Psi }\beta
\partial _{\tau }\partial _{z}\Psi +\frac{1}{\gamma ^{2}}\frac{1}{\Psi }%
\partial _{z}^{2}\Psi -G_{1}^{\prime }\left( \Psi ^{2}\right) \right) \zeta
^{2}.  \label{ReQ2}
\end{equation}%
By (\ref{logb}) with $\Psi ^{2}=\mathrm{e}^{2\sigma }\Psi _{1}^{2}$, we
obtain that 
\begin{equation*}
G_{1}^{\prime }\left( \mathrm{e}^{2\sigma }\Psi _{1}^{2}\right) =-2\sigma
+G_{1}^{\prime }\left( \Psi _{1}^{2}\right) .
\end{equation*}%
Obviously, $\Psi _{1}$ does not depend on $\sigma $, and $\Psi _{1}$ has the
form( \ref{psigam}), but with $\sigma =0$ in the definition of $\mathring{%
\psi}_{1}$ in (\ref{psi1sig}). \ Such a $\mathring{\psi}_{1}$ satisfies the
equation 
\begin{equation}
\partial _{3}^{2}\mathring{\psi}_{1}-G_{a1}^{\prime }\left( \mathring{\psi}%
_{1}^{\ast }\mathring{\psi}_{1}\right) \mathring{\psi}_{1}=0,  \label{G1D}
\end{equation}%
and from (\ref{ReQ2}) we obtain that%
\begin{equation}
Q=-\frac{\zeta ^{2}}{\Psi }\partial _{\tau }^{2}\Psi +\frac{\zeta ^{2}}{\Psi 
}\partial _{\tau }\beta \partial _{z}\Psi +\frac{2\zeta ^{2}}{\Psi }\beta
\partial _{\tau }\partial _{z}\Psi +\frac{\zeta ^{2}}{\Psi }\left( \frac{1}{%
\gamma ^{2}}-\frac{1}{\gamma _{0}^{2}}\right) \partial _{z}^{2}\Psi +2\zeta
^{2}\sigma .  \label{ReQ1}
\end{equation}%
Now we rewrite the complex equation (\ref{Qeq}) as a system of two real
equations. The real part of (\ref{Qeq}) divided by $\Psi $ yields the
following quadratic equation 
\begin{equation}
Q-2\gamma \Phi +\Phi ^{2}-2\beta \gamma Z-Z^{2}=0.  \label{ReQeq}
\end{equation}%
The small solution $Z$ to this equation is given by the formula 
\begin{equation}
Z=\Theta \left( \tau ,\Phi \right) =-\beta \gamma +\left( \Phi ^{2}-2\gamma
\Phi +\beta ^{2}\gamma ^{2}+Q\right) ^{1/2}\frac{\beta }{\left\vert \beta
\right\vert }.  \label{ZfiQ}
\end{equation}%
The imaginary part of (\ref{Qeq}) divided by $\zeta \Psi $ yields 
\begin{equation}
-2\Phi \left( \partial _{\tau }-\beta \partial _{z}\right) \ln \Psi -\left(
\partial _{\tau }-\beta \partial _{z}\right) \Phi +\partial _{z}Z+2Z\partial
_{z}\ln \Psi =0,  \label{ImQeq0}
\end{equation}%
the coefficients are expressed in terms of the given $\Psi $ defined by (\ref%
{psigam}), (\ref{psi1sig}), and can be written explicitly: 
\begin{equation}
\partial _{z}\ln \Psi =-\gamma _{0}^{2}z,\text{\quad }\partial _{\tau }\ln
\Psi =\partial _{\tau }\sigma ,\text{\quad }\gamma _{0}=\left( 1-\frac{1}{%
\mathrm{c}^{2}}v^{2}\left( 0\right) \right) ^{-1/2}.  \label{dlog}
\end{equation}%
To determine a \emph{small} solution $\Phi $ of (\ref{ImQeq0}), (\ref{ZfiQ})
in the strip $\Xi $ we impose the condition%
\begin{equation}
\Phi =0\text{ if\ }z=0,\text{\quad }-\infty <\tau <\infty .  \label{Zy0}
\end{equation}%
The exact solution $\Phi $ of the equation (\ref{ImQeq0}), where $Z\ =\Theta
\left( \tau ,\Phi \right) $ satisfies (\ref{ZfiQ}),\ is a solution of the
following quasilinear first-order equation 
\begin{equation}
\partial _{\tau }\Phi -\beta \mathbf{\partial }_{z}\Phi -\Theta _{\Phi
}\left( \tau ,\Phi \right) \mathbf{\partial }_{z}\Phi =-2\Phi \zeta \left(
\partial _{\tau }-\beta \partial _{z}\right) \ln \Psi +2\Theta \partial
_{z}\ln \Psi +\Theta _{z},  \label{Zfieq2}
\end{equation}%
where $\Theta _{\Phi }$ and $\Theta _{z}$\ are partial derivatives of $%
\Theta \left( \tau ,\Phi \right) $ ($\Theta $ depends on $z$ via $Q=\func{Re}%
Q$) with $\Phi $ satisfying condition (\ref{Zy0}). According to the method
of characteristics, we write based on (\ref{Zfieq2}) the following equations
for the characteristics:\ 
\begin{gather}
\frac{d\tau }{ds}=1,\quad \frac{dz}{ds}=-\Theta _{\Phi }\left( \tau ,\Phi
\right) -\beta \left( \tau \right) ,  \label{exLC1} \\
\frac{d\Phi }{ds}=-2\Phi \left( \partial _{\tau }-\beta \partial _{z}\right)
\ln \Psi +2\Theta \left( \tau ,\Phi \right) \partial _{z}\ln \Psi +\Theta
_{z},  \label{exLC3}
\end{gather}%
with the initial data derived from (\ref{Zy0}) on the line $z=0$: 
\begin{equation}
\tau _{s=0}=\tau _{0},\quad z_{s=0}=0,\quad \Phi _{s=0}=0.  \label{LCin}
\end{equation}%
Note that $\Psi =\pi ^{-1/4}\mathrm{e}^{-\gamma _{0}^{2}z^{2}/2+\sigma }$ is
a given function of $\tau ,z$,\ and $\beta ,\gamma $ are given functions of $%
\tau .$ By the method of characteristics, the solution of (\ref{Zfieq2}), (%
\ref{Zy0}) can be found by integration along the integral curves. If the
right-hand side in (\ref{exLC3}) vanishes at $\Phi =0,$ the exact solution $%
\Phi \ $would be zero along the curve for every $\tau _{0}$;\ hence it would
be zero in $\Xi $. Therefore the magnitude of $\Phi $ on the curve in the
strip $\Xi $ is determined by the magnitude of the remaining term, 
\begin{equation}
F_{1}=2\Theta \left( \tau ,0\right) \partial _{z}\ln \Psi +\Theta _{z}\left(
\tau ,0\right) ,\   \label{F1eq}
\end{equation}%
in the right-hand side. The system (\ref{exLC1})-(\ref{LCin}) is still too
complex to hope for a closed form solution, but here we want only to show
that this solution is small; this implies, in turn, that the potential $%
\varphi _{2}$ is small as well. To this end, we show below that (\ref{exLC1}%
)-(\ref{exLC3}) is a small perturbation of a simpler system which
corresponds to the principal part of (\ref{Zfieq2}) for small $\zeta $.

\subsection{Small auxiliary phases and small balancing potential}

Now we estimate the leading part of $\Phi $ for small $\zeta $. We obtain
from (\ref{ReQ1}) the estimate%
\begin{equation}
\left\vert Q\right\vert +\left\vert \partial _{z}Q\right\vert +\left\vert
\partial _{\tau }Q\right\vert =O\left( \epsilon \zeta ^{2}+\epsilon
_{1}\zeta ^{2}\right) .  \label{ReQ2o}
\end{equation}%
Obviously, if $Q=0$ then $\Theta \left( \tau ,0\right) =0$. Since in the
formula (\ref{ZfiQ}) $\beta ^{2}\gamma ^{2}>0,$ this formula determines for
small $\zeta $\ a smooth function $\Theta $ of small $\Phi \left( \tau
,z\right) $\ in the strip $\Xi \left( \theta \right) ,$ and one can see from
(\ref{ReQ1}) that the expression for $Q$ is a quadratic polynomial in $z$
with the common factor $\zeta ^{2}$ and bounded coefficients. Therefore $%
\Theta \left( \tau ,\Phi \right) $ is a regular function of \emph{small} $%
\Phi $ in the strip $\Xi \left( \theta \right) $ as long as 
\begin{equation}
\zeta ^{2}\theta ^{2}+\zeta ^{2}<<1.  \label{zetthet}
\end{equation}%
\ Hence we can take 
\begin{equation}
\theta \rightarrow \infty \text{ \ \ \ as \ }\zeta \rightarrow 0,
\label{thetinf}
\end{equation}%
and, therefore, the width $\theta $ of the strip where we have the exact
solution of (\ref{Zfieq2}), (\ref{ZfiQ}) is arbitrary large. If $\Phi
=O\left( \zeta ^{2}\epsilon +\zeta ^{2}\epsilon _{1}\right) $ (certainly
this assumption is true for small $\left\vert z\right\vert $ thanks to (\ref%
{Zy0}), the argument below shows that we can extend this estimate to the
entire strip $\Xi \left( \theta \right) $) we have 
\begin{equation}
Z=\Theta \left( \tau ,\Phi \right) =-\frac{1}{\beta }\Phi +O\left( \zeta
^{2}\epsilon +\epsilon _{1}\zeta ^{2}\right) .  \label{ZfiQ1}
\end{equation}%
This implies the following estimate of $F_{1}$ \ in (\ref{F1eq})%
\begin{equation*}
F_{1}=O\left( \zeta ^{2}\epsilon +\zeta ^{2}\epsilon _{1}\right) .
\end{equation*}%
If we replace $\Theta $ by its principal term from (\ref{ZfiQ1}), we obtain
from (\ref{exLC1})-(\ref{exLC3}) a simpler system for the resulting
approximation $\mathring{\Phi}$: 
\begin{gather}
\frac{d\tau }{ds}=1,\quad \frac{dz}{ds}=-\beta +\frac{1}{\beta },
\label{exLC10} \\
\frac{d\mathring{\Phi}}{ds}=-2\mathring{\Phi}\left( \partial _{\tau }-\beta
\partial _{z}\right) \ln \Psi -\frac{2}{\beta }\mathring{\Phi}\partial
_{z}\ln \Psi ,  \label{exLC31}
\end{gather}%
A general solution for this system can be written explicitly. Namely, the
characteristic curves are given by 
\begin{equation}
\tau =\tau _{0}+s,\quad z=z\left( \tau ,\tau _{0}\right) =\int_{\tau
_{0}}^{\tau }\frac{1}{\beta }\left( 1-\beta ^{2}\right) d\tau ,  \label{ztau}
\end{equation}%
and the value of the phase $\mathring{\Phi}$ is given by the formula%
\begin{equation}
\mathring{\Phi}\left( \tau ,z\right) =\mathring{\Phi}\left( \tau
_{0},0\right) \frac{\Psi ^{2}\left( \tau _{0},0\right) }{\Psi ^{2}\left(
\tau ,z\right) }.  \label{Fitau}
\end{equation}%
The integral curves $\tau ,z\left( \tau ,\tau _{0}\right) $\ obviously cover
the entire $\left( \tau ,z\right) $-plane. Since the original problem (\ref%
{exLC1})-(\ref{exLC3}) is a perturbation of the above system of order $\zeta
^{2}\epsilon +\zeta ^{2}\epsilon _{1}$, the nice properties of (\ref{exLC10}%
), (\ref{exLC31}) imply that\ the characteristic equations (\ref{exLC1})-(%
\ref{exLC3}) and their integral curves also have nice properties, and $\Phi $
and $\mathring{\Phi}$ are close one to another. According to (\ref{LCin}) we
set $\mathring{\Phi}\left( \tau _{0},0\right) =0$, hence $\mathring{\Phi}%
\left( \tau ,z\right) =0$, and the exact solution $\Phi $ of (\ref{Zfieq2}),
(\ref{Zy0})\ in the strip $\Xi $ is of order $\zeta ^{2}\epsilon +\zeta
^{2}\epsilon _{1}$ for small $\zeta $.

Let us verify now that the balancing potential\ is small. The solution of (%
\ref{Zfieq2}), (\ref{Zy0}), obtained by integrating (\ref{exLC1})-(\ref{LCin}%
), satisfies in $\Xi $ estimates%
\begin{equation}
\left\vert \Phi \right\vert =O\left( \left( \zeta ^{2}\epsilon +\zeta
^{2}\epsilon _{1}\right) \left\vert z\right\vert \right) ,\quad \left\vert
\partial _{z}\Phi \right\vert +\left\vert \partial _{z}^{2}\Phi \right\vert
=O\left( \zeta ^{2}\epsilon +\zeta ^{2}\epsilon _{1}\right) .  \label{Fieps2}
\end{equation}%
From (\ref{denZ})-(\ref{denZ2}) we obtain that 
\begin{equation}
\varphi _{2}=\frac{m\mathrm{c}^{2}}{q}O\left( \zeta ^{2}\epsilon \left\vert
z\right\vert +\zeta ^{2}\epsilon _{1}\left\vert z\right\vert \right) ,
\label{fi2small}
\end{equation}%
and obviously the balancing potential vanishes when $\zeta \rightarrow 0$.

Let us show now that the constructed potential $\varphi _{\mathrm{ex}%
}=\varphi _{\mathrm{ac}}+\varphi _{2}$ satisfies the asymptotic local
homogeneity\emph{\ }condition (\ref{aac0}). Definition (\ref{onerf}) takes
the form 
\begin{equation}
\frac{1}{R_{f}}=\frac{\max_{\left\vert x-r\right\vert \leq \theta a}\left(
\left\vert \partial _{x}^{2}\varphi _{\mathrm{ex}}\right\vert \right) }{%
\max_{t}\left\vert \partial _{x}\varphi _{\mathrm{ac}}\right\vert }=\frac{%
\max_{\left\vert z\right\vert \leq \theta }\left( \left\vert \partial
_{z}^{2}\varphi _{\mathrm{ex}}\right\vert \right) }{a\max_{\tau }\left\vert
\partial _{z}\varphi _{\mathrm{ac}}\right\vert }.  \label{onerf1D}
\end{equation}%
To express the magnitude of the gradient of the potential $\varphi _{\mathrm{%
ac}}$ in terms of the parameters corresponding to the mild accelerations, we
use relations (\ref{Newt}) and (\ref{betep0}): 
\begin{equation*}
\partial _{z}\varphi _{\mathrm{ac}}=a\partial _{y}\varphi _{\mathrm{ac}}=-a%
\frac{m}{q}\partial _{t}\left( \gamma v\right) =-\frac{m\mathrm{c}^{2}}{q}%
\partial _{\tau }\left( \gamma \beta \right) ,
\end{equation*}%
resulting in the estimate%
\begin{equation}
\max_{\tau }\left\vert \partial _{z}\varphi _{\mathrm{ac}}\right\vert =\frac{%
m\mathrm{c}^{2}}{q}O\left( \epsilon \right) .  \label{dzac}
\end{equation}%
\ Using (\ref{denZ}), (\ref{denZ2}) and taking into account the definition
of $a_{\mathrm{C}}$, we obtain 
\begin{equation}
\mathbf{\partial }_{z}^{2}\Phi +\mathbf{\partial }_{\tau }\mathbf{\partial }%
_{z}Z-\beta \mathbf{\partial }_{z}^{2}Z=\frac{q}{m\mathrm{c}^{2}}\mathbf{%
\partial }_{z}^{2}\varphi _{2}.  \label{Fizfi2}
\end{equation}%
We observe that, since $\varphi _{\mathrm{ac}}$ is linear, $\partial
_{z}^{2}\varphi _{\mathrm{ex}}=\partial _{z}^{2}\varphi _{2},$ and using (%
\ref{Fieps2}) and (\ref{ZfiQ1}), \ we conclude that in the strip $\Xi $%
\begin{equation*}
\mathbf{\partial }_{z}^{2}\varphi _{\mathrm{ex}}\ =\frac{m\mathrm{c}^{2}}{q}%
O\left( \zeta ^{2}\epsilon +\epsilon _{1}\zeta ^{2}\right) .
\end{equation*}%
Consequently, formula (\ref{onerf1D}) takes the form 
\begin{equation}
\frac{1}{R_{f}}=\frac{O\left( \zeta ^{2}\epsilon +\epsilon _{1}\zeta
^{2}\right) }{aO\left( \epsilon \right) }=\frac{1}{a}O\left( \zeta ^{2}+%
\frac{\epsilon _{1}}{\epsilon }\zeta ^{2}\right) .  \label{Rf1D}
\end{equation}%
We obtain that (\ref{aac0}) holds in the asymptotic regime 
\begin{equation}
\zeta \rightarrow 0,\quad \frac{\epsilon _{1}}{\epsilon }\zeta
^{2}\rightarrow 0,  \label{asreg}
\end{equation}%
which is consistent with the assumptions (\ref{epzetep}), (\ref{zetzet0}). \
The above analysis shows also that the magnitude of the balancing potential $%
\varphi _{2}$ is vanishingly small compared with the magnitude of the
accelerating potential.

The summary of the above argument is that we can find a potential $\varphi _{%
\mathrm{ex}}\left( t,x\right) =\varphi _{\mathrm{ac}}+\varphi _{2}$\ such
that (i) the function $\psi ^{\prime }$ with the Gaussian profile defined by
(\ref{psipsi}) is an exact solution of the KG equation (\ref{KGy}) in a
widening strip $\Xi \left( \theta \right) $; (ii) the balancing potential $%
\varphi _{2}\left( t,x\right) $ vanishes asymptotically, and the relative
magnitude of the balancing potential $\varphi _{2}\left( t,x\right) $
compared with the accelerating potential $\varphi _{\mathrm{ac}}$ tends to
zero as $\zeta \rightarrow 0$. The charge distribution $\psi $ provides an
example of a localized solution in its strongest form, namely, an
accelerating solution with a fixed Gaussian shape. The possibility to
preserve localization which we discuss in this section concerns microscopic
details of the charge evolution, and naturally the ratio $\zeta =a_{\mathrm{C%
}}/a$ of the charge size to the Compton wavelength $a_{\mathrm{C}}$ plays an
important role. In the context of Section \ref{secCL}, $a$ is vanishingly
small compared with the macroscopic scale $R_{f}$ as (\ref{Rf1D}) clearly
shows.

\textbf{Acknowledgment.} The research was supported through Dr. A. Nachman
of the U.S. Air Force Office of Scientific Research (AFOSR), under grant
number FA9550-11-1-0163.

\end{document}